\documentclass{aa}  
\usepackage{graphicx}
\usepackage{txfonts}
\usepackage{xcolor}
\bibpunct{(}{)}{;}{a}{}{,} 

\begin{document} 
\title{Companions to \textit{Kepler} giant stars: A long-period eccentric 
substellar companion to KIC~3526061 and a stellar companion to HD~187878}


\author{Marie Karjalainen\inst{1}\thanks{These authors contributed equally to this work.}
       \and Raine Karjalainen\inst{1}$^{\star}$
       \and Artie P.~Hatzes\inst{2} \and Holger Lehmann\inst{2}
       \and Pierre Kervella\inst{3} \and Saskia Hekker\inst{4,5,6} 
       \and Hans Van~Winckel\inst{7} \and Jakub \"{U}berlauer\inst{8}
       \and Michaela V\'{i}tkov\'{a}\inst{9} \and Marek Skarka\inst{1,9}
       \and Petr Kab\'{a}th\inst{1} \and Saskia Prins\inst{7} 
       \and Andrew Tkachenko\inst{7}
       \and William D.~Cochran\inst{10}
       \and Alain Jorissen\inst{11}
       }

\institute{Astronomical Institute, Czech Academy of Sciences, 251 65 Ond\v{r}ejov, Czech Republic\\
         \email{marie.karjalainen@asu.cas.cz}
         \and
         Th\"uringer Landessternwarte Tautenburg, Sternwarte 5, Tautenburg, D-07778, Germany
         \and LESIA, Observatoire de Paris, Universit\'{e} PSL, CNRS, Sorbonne
         Universit\'{e}, Univ. Paris Diderot, Sorbonne Paris Cit\'{e}, 5 place Jules Janssen, 
         92195 Meudon, France
         \and
         Max-Planck-Institut f\"ur Sonnensystemforschung, Justus-von-Liebig-Weg 3, 
         G\"ottingen, D-37077, Germany
         \and
         Center for Astronomy (ZAH/LSW), Heidelberg University, K\"{o}nigstuhl 12, 
         69117 Heidelberg, Germany
         \and
         Heidelberg Institute for Theoretical Studies (HITS) gGmbH, 
         Schloss-Wolfsbrunnenweg 35, 69118 Heidelberg, Germany
         \and
         Instituut voor Sterrenkunde, KU Leuven, Celestijnenlaan 200D bus 2401, Leuven, 3001, Belgium         
         \and
         Astronomical Institute, Faculty of Mathematics and Physics, Charles University, V
         Hole\v{s}ovi\v{c}k\'{a}ch 2, 18000 Praha 8, Czech Republic
         \and
         Department of Theoretical Physics and Astrophysics, Masaryk University,
         Kotl\'{a}\v{r}sk\'{a} 2, 61137 Brno, Czech Republic
         \and
         McDonald Observatory and Center for Planetary Systems Habitability, The University of Texas,
         Austin, Texas, USA
         \and
         Institut d\textquoteright Astronomie et d\textquoteright Astrophysique, 
         Universit\'e Libre de Bruxelles C.P. 226, 
         Boulevard du Triomphe, 1050 Bruxelles, Belgium
         }
         
\date{Received September 15, 1996; accepted March 16, 1997}

\abstract
{Our knowledge of populations and occurrence of planets orbiting evolved 
intermediate-mass stars is still incomplete. In 2010 we 
started a planet-search program among 95 giant stars 
observed by the \textit{Kepler} mission to increase the sample 
of giant stars with planets and with reliable estimates of stellar masses
and radii.} 
{We present the two systems KIC~3526061 and HD~187878
from our planet-search program for which we could characterise their
companions.} 
{We used precise stellar radial velocity measurements taken with four
different echelle spectrographs to derive an orbital solution. 
We used \textit{Gaia} astrometric measurements to obtain the inclination
of the HD~187878 system and \textit{Kepler} photometric observations to estimate the stellar
mass and radius.}
{We report the discovery of a sub-stellar and a stellar companion around two
intermediate-mass red giant branch stars. KIC~3526061~b is most likely a brown dwarf with a 
minimum mass of $18.15\pm 0.44\,{\rm{M_{Jupiter}}}$ in a long-period
eccentric orbit, with the orbital period $3552_{-135}^{+158}$ d and 
orbital eccentricity $e=0.85\pm 0.01$. It is the most evolved system found 
having a sub-stellar companion with such a large eccentricity and wide separation. 
HD~187878~B has a minimum mass of $78.4\pm 2.0\,{\rm{M_{Jupiter}}}$. 
Combining the spectroscopic orbital parameters with the astrometric proper 
motion anomaly we derived an orbital inclination $i=9.8^{+0.4}_{-0.6}$ deg, which 
corresponds to the companion's mass in the stellar regime of 
$0.51^{+0.04}_{-0.02}\,\rm{M_{\odot}}$.}
{A sub-stellar companion of KIC~3526061 extends the sample of known red giant
branch stars with sub-stellar companions on very eccentric wide orbits and
might provide a probe of the dynamical evolution of such systems over time.}

\keywords{methods: observational -- 
          techniques: radial velocities --
          techniques: spectroscopic -- 
          Stars: individual: KIC~3526061 --
          Stars: individual: HD~187878 -- 
          brown dwarfs
          }

\titlerunning{Companions to \textit{Kepler} giant stars KIC~3526061 and
HD~187878}

\maketitle
%

\section{Introduction}

The two most successful techniques for the detection of exoplanets are the transit and Doppler 
methods and both are biased towards host stars with masses less than
$\sim1\,\rm{M_{\odot}}$. Detecting transit signals due to
planets for intermediate-mass stars (1.2 -- 2 $\rm{M_{\odot}}$) is more
difficult than for lower-mass dwarf stars due to a larger stellar radius of
more massive stars. Likewise, intermediate-mass main sequence stars
are ill-suited for the radial velocity (RV) method. They
are hot and thus have few spectral lines for the RV measurement. 
More problematical, the stars have appreciable amounts of rotation which broadens the spectral 
lines and makes them more shallow which further decreases the RV precision
making the detection of planetary companions difficult.

On the other hand, intermediate-mass stars that have evolved to giant stars are cool 
so they have more spectral lines and usually rotate slowly which means that
spectral lines are narrower. One can easily 
obtain an RV precision of several $\rm{m\,s^{-1}}$ on a 2 $\rm{M_{\odot}}$ giant star compared 
to tens of $\rm{m\,s^{-1}}$ for its main sequence progenitor. The giant stars thus 
serve as proxies for planet searches around intermediate-mass early-type main sequence stars. 

Since the discovery of first exoplanets around K-giant stars \citep{Hatzes93, Frink02},   
several teams have been surveying so-called retired A stars using a RV
method with a goal to learn more about the occurrence and properties of planets 
around more massive stars. These surveys discovered most of the over 150 exoplanets 
orbiting giant stars, which is 3 \% of the total exoplanets known to 
date\footnote{http://exoplanet.eu/}. Our list of exoplanets
orbiting giant stars excludes those orbiting subgiants, and includes
companions with the mass or the minimum mass of $< 13\,\rm{M_{Jupiter}}$.

Studying extrasolar planets around K-giant stars is essential to find out   
the planetary occurrence versus the stellar mass and metallicity, which in
turn is important to better understand the planet-formation and evolution processes. 
The disk instability model predicts that there should be no dependence of
stellar host mass on
planet formation and physical stellar properties \citep{Boss06}. On the other hand, the core-accretion
formation process, which is believed to be at the origin of most of the planets, predicts an increase 
in giant planet frequency with stellar mass up to
3 $M_{\odot}$ \citep{Kennedy08}. The majority of studies confirms the trend of the higher 
giant planet occurrence for higher stellar host masses \citep[for example][]{Bowler10, Johnson10, 
Reffert15, Jones16, Ghezzi18},
which provides an additional support for the core accretion mechanism of planet formation. 
However, unlike for a main sequence star it is more problematic 
to determine the stellar mass of a giant star. Evolutionary tracks
for stars covering a wide range of masses all converge to the
similar region of the H-R diagram. 

Fortunately, the stellar mass can be derived from solar-like oscillations 
that are ubiquitous in K giant stars. 
Their first firm discovery in a giant star was made by \citet{Frandsen02}. However, only later they  
were unambiguously found in late-type giant stars, due to the \textit{CoRoT}
\citep{DeRidder09} and the \textit{Kepler} \citep{Huber10, Kallinger10} space missions.
The \textit{Kepler} mission was monitoring a sample of over 13,000
red-giant stars which can be used for asteroseismic studies and 
have been analysed to determine their fundamental stellar parameters
\citep{Stello13}.

To increase the sample of giant stars with planets and to have more complete information 
of individual systems, we started a planet-search program among \textit{Kepler} 
asteroseismic-giant stars in 2010. Our advantage over other ground-based RV surveys of
evolved stars is that we can determine reliable stellar properties such as the
stellar mass and radius via the asteroseismic analysis, using the \textit{Kepler} 
(Borucki et al. 2010) photometric data. These are characteristics well known
so far only for nine planet-hosting giant stars observed by space missions,
which were found to have transiting planets
\citep{Huber13,Lillo-Box14,Ciceri15,Ortiz15,Quinn15,Grunblatt16,Grunblatt17,Jofre20,Grunblatt22}.
Our sample of \textit{Kepler} giant stars contains 95 targets, which is a statistically 
significant number given an expected giant-planet occurrence of
$\sim$17 \% \citep{Ghezzi18} for the mean stellar mass of
our sample in the range 1.5 -- 2 $\rm{M_{\odot}}$.
We have chosen our targets from an initial list of giant stars from the {\it Kepler} 
Input Catalogue, which we ordered in brightness. Based on the {\it Kepler} photometry 
we removed binaries and targets that were clearly not giant stars. Then we
selected our sample from the brightest targets down to V=10.74 mag and  
distributed it over four different telescopes in order to maximise the detection and to 
minimise the impact of telescope resources at a single site. 
The goal of our planet-search program is to characterise each individual target and conclude 
on the existence of sub-stellar companions. We have published the discovery of a
planetary candidate around the evolved low-mass \textit{Kepler} giant star 
HD~175370 \citep{Hrudkova17}. Here, we report the discovery of another two
companions to \textit{Kepler} giant stars from our planet-search program,
for which we could conclude on the nature of companions. We have found
a brown dwarf candidate orbiting the red giant branch star KIC~3526061 and a
stellar mass companion orbiting the red giant branch star HD~187878. 
KIC~3526061 is the most evolved system found so far having a sub-stellar
companion with such a large eccentricity and wide separation. 

This paper is organised as follows. In Sect.~\ref{observations}, we give
an overview of our observations of KIC~3526061 and HD~187878, and describe 
the four different echelle spectrographs we have used in this study. 
Section~\ref{KIC3526061} is devoted to KIC~3526061, where we derive stellar 
parameters and orbital solution, analyse stellar activity and discuss our
results, particularly the origin of a large orbital eccentricity.
In a similar manner, we describe our analysis and results for HD~187878 in
Sect.~\ref{HD187878}.
In Sect.~\ref{Conclusions}, we provide a brief summary and conclusions.  

\section{Observations and data reduction}\label{observations}

We have observed KIC~3526061 since August 2012 using the Robert G. Tull 
Coud\'e cross-dispersed echelle spectrograph (TS2)
of the 2.7-m Harlan J. Smith Telescope at the McDonald Observatory in Texas,
U.S. We obtained 17 spectra with a S/N of $\sim$ 100 per pixel in the extracted 
spectrum. Since April 2018 we have monitored this star also using the fibre-fed
High Efficiency and Resolution Mercator Echelle Spectrograph (HERMES) 
at the 1.2-m Mercator Telescope on La Palma, Canary Islands, Spain.
We obtained eight spectra with a S/N of $\sim$ 80 per pixel in 
the extracted spectrum. The RV measurements from both sites are listed in
Table~\ref{kic_rvs}.

We have observed HD~187878 since March 2010 using the coud\'e echelle
spectrograph at the 2-m Alfred Jensch Telescope at the Th\"uringer
Landessternwarte Tautenburg (TLS), Germany.
We obtained 54 spectra with a S/N of $\sim$ 100 per pixel in the extracted 
spectrum. Since July 2011 we have monitored this star using the HERMES spectrograph. 
We obtained 25 spectra with a S/N of $\sim$ 85 per pixel in the extracted spectrum. 
Since September 2020 we have observed this star with the Ond\v{r}ejov
Echelle Spectrograph (OES) installed at the 2-m Perek telescope at the Astronomical 
Institute of the Czech Academy of Sciences in Ond\v{r}ejov, Czech Republic. 
We obtained 21 spectra with a S/N of $\sim$ 30 -- 180 per pixel in the extracted spectrum.
The RV measurements from all sites are listed in Table~\ref{hd_rvs}.

\begin{table}
\caption{RV measurements of KIC~3526061. At the McDonald Observatory
we used an iodine cell which resulted to RVs relative to a stellar template,
while from the Mercator telescope RVs are absolute. 
RVs were corrected for the barycentre motion, using the program
\textsc{BarCor} (https://stelweb.asu.cas.cz/$\sim$marie/Barcor/). 
The RV uncertainties correspond to instrumental errors.
This table is available in its entirety in machine-readable 
form.}\label{kic_rvs}
\centering
\begin{tabular}{lrr}\hline
BJD & RV & $\sigma_{\rm{RV}}$\\
(d) & ($\rm{m\,s^{-1}}$) & ($\rm{m\,s^{-1}}$)\\ 
\hline
\noalign{\smallskip}
\multicolumn{3}{l}{McDonald}\\
\noalign{\smallskip}
2456168.845738 &   -55.7 & 20.4 \\
2456550.715116 &   -20.1 & 14.4 \\
\multicolumn{3}{l}{\dots}\\
\noalign{\smallskip}
\multicolumn{3}{l}{HERMES}\\
\noalign{\smallskip}
2458218.683338 &-27465.0 &  5.0 \\
2458261.513262 &-27419.0 &  3.0 \\
\multicolumn{3}{l}{\dots}\\
\noalign{\smallskip}         
\hline
\end{tabular}
\end{table} 

\begin{table}
\caption{RV measurements of HD~187878 from three different sites.
At TLS and Ond\v{r}ejov we used an iodine cell which resulted in  
RVs relative to a stellar template, while with HERMES we got absolute RVs.
RVs were corrected for the barycentre motion. 
The RV uncertainties correspond to instrumental errors. The difference
between TLS new and old is explained in the text. This table is available in 
its entirety in machine-readable form.}\label{hd_rvs}
\centering
\begin{tabular}{lrr}\hline
BJD & RV & $\sigma_{\rm{RV}}$\\
(d) & ($\rm{m\,s^{-1}}$) & ($\rm{m\,s^{-1}}$)\\ 
\hline
\noalign{\smallskip}             
\multicolumn{3}{l}{HERMES}\\
\noalign{\smallskip}             
2455745.529925 & -18026  &   2 \\
2455771.640696 & -18121  &   2 \\
\multicolumn{3}{l}{\dots}\\
\noalign{\smallskip}         
\multicolumn{3}{l}{TLS old}\\
\noalign{\smallskip}
2455279.578207 &    390.8 &   8.2  \\
2455357.394732 &    443.7 &   9.1  \\
\multicolumn{3}{l}{...}\\
\noalign{\smallskip}         
\multicolumn{3}{l}{TLS new}\\
\noalign{\smallskip}         
2456783.548644 &    470.0 &  17.1 \\
2456797.355630 &    456.0 &  13.7 \\
\multicolumn{3}{l}{...}\\
\noalign{\smallskip}         
\multicolumn{3}{l}{Ond\v{r}ejov}\\
\noalign{\smallskip}         
2459108.399065 &  -496.5 &  24.5 \\
2459124.276912 &  -494.2 &  28.9 \\
\multicolumn{3}{l}{...}\\
\noalign{\smallskip}         
\hline
\end{tabular}
\end{table} 

\subsection{TLS data}

The coud\'e echelle spectrograph at TLS provides 
a wavelength range of 4670 -- 7400~\AA~and a spectral resolving power of
67,000. We have reduced the data using standard \textsc{IRAF}\footnote{The Image
Reduction and Analysis Facility (\textsc{IRAF}) is 
distributed by the National Optical Astronomy Observatories, which are
operated by the Association of Universities for Research in Astronomy, Inc., under
cooperative agreement with the National Science Foundation.} procedures
(bias subtraction, flat-field correction, extraction of individual echelle
orders, wavelength calibration, subtraction of scattered light, cosmic rays removal and
spectrum normalisation). We have used an iodine absorption cell placed in the optical path 
just before the slit of the spectrograph as the wavelength reference.  
The calculation of the RVs largely followed the method outlined in \citet{Valenti95},
\citet{Butler96} and \citet{Endl00}, and takes into account changes in the instrumental
profile. We note that the measured RVs are relative to a stellar template which is 
an iodine-free spectrum and are not absolute values.
Since May 2014 a new echelle grating has been mounted at the coud\'e
spectrograph at TLS. With the new set-up it was necessary to treat data as
an independent data set, using a new stellar template as a reference. Throughout the
paper, we refer to data taken before this change as TLS old data and data
taken after the change as TLS new data.

\subsection{McDonald data}

The TS2 coud\'e echelle spectrograph at the McDonald Observatory provides a 
wavelength range of 3400 -- 10,900~\AA~and a spectral resolving power of
60,000. We have reduced the data using standard \textsc{IRAF} procedures
and used an iodine absorption cell as the wavelength reference, 
similarly as for the TLS data. More details about the TS2 spectrograph 
can be found in \citet{Tull95}.

\subsection{HERMES data}

For the HERMES spectrograph, we used a simultaneous ThArNe wavelength reference mode 
in order to achieve as accurate RV measurements as possible. The wavelength range of
HERMES is 3770 -- 9000~\AA. 
Before August 2018 we used the lower-resolution mode of HERMES, which provided
us with a spectral resolving power of 62,000. Since August 2018 an option of
using a high-resolution fibre (HRF) mode with a simultaneous ThArNe wavelength
reference became available, so we started to observe with a HRF mode
providing a spectral resolving power of 85,000.    
More details about the HERMES spectrograph can be found in
\citet{Raskin11}. We have used a dedicated automated data reduction pipeline
and RV toolkit (\textsc{HermesDRS}) to reduce the data and calculate
absolute RVs. The spectral mask of Arcturus on 
the velocity scale of the IAU RV standards was used for the cross-correlation.

\subsection{Ond\v{r}ejov data}

The OES spectrograph in Ond\v{r}ejov provides a wavelength range of 
3750 -- 9200~\AA~and a spectral resolving power of 50,000. More details about the 
OES spectrograph can be found in \citet{Koubsky04} and \citet{Kabath20}.
We have reduced the data using standard \textsc{IRAF} procedures,
and used an iodine absorption cell as the wavelength reference. RVs were calculated using the 
Velocity and Instrument Profile EstimatoR
(\textsc{viper})\footnote{https://github.com/mzechmeister/viper}
\citep{Zechmeister21}, which is a \textsc{python}-based software to calculate RVs of 
stellar spectra taken using iodine cell or other gas cells. 

\section{KIC~3526061}\label{KIC3526061}

\subsection{Stellar properties}\label{properties_kic}

KIC~3526061 has a visual magnitude of $m_V=10.37\pm 0.04$ mag \citep{Hog00}. 
The parallax was determined from \textit{Gaia} EDR3 data
as $2.504\pm 0.012$ mas \citep{Gaia16, Gaia21a}, which implies an absolute magnitude 
$M_V=2.36\pm 0.04$ mag. Table~\ref{kic_stellar_par} lists stellar parameters 
known from literature together with those determined in this work. 

\begin{table*}\tabcolsep=2.7pt
\caption{Stellar parameters of KIC~3526061 from this work together with
those known from literature.} \label{kic_stellar_par}
\centering
\begin{tabular}{lllclll}
\hline 
\noalign{\smallskip}
Parameter & Value & Reference & & Parameter & Value & Reference\\
\noalign{\smallskip}
\hline 
\noalign{\smallskip}
$m_V$ (mag) & $10.37\pm 0.04$ & \citet{Hog00} &                                     $\quad\quad$ & $\Delta\nu$ ($\mu$Hz) & $10.7\pm 0.3$ & \citet{Hekker11}\tablefootmark{a}\\                             
$B-V$ (mag) & $0.96\pm 0.08$ & \citet{Hog00} &                                      &$\Delta\nu$ ($\mu$Hz) & $10.73\pm 0.05$ & \citet{Mosser14}\\
Parallax (mas) & $2.5042\pm 0.0123$ & Gaia Coll. et al. (2021a) &                   &$\Delta\nu$ ($\mu$Hz) & $10.64\pm 0.23$ & \citet{Pinsonneault14}\\
$M_V$ (mag) & $2.36\pm 0.04$ & This work &                                          &$\Delta\nu$ ($\mu$Hz) & $10.670\pm 0.014$ & \citet{Huber17}\tablefootmark{b}\\
Distance (pc) & $399\pm 2$ & This work &                                            &$\Delta\nu$ ($\mu$Hz) & $10.667\pm 0.004$ & \citet{Pinsonneault18}\\
Distance (pc) & $412^{+10}_{-18}$ & \citet{Rodrigues14} &                           &$\Delta\nu$ ($\mu$Hz) & 10.71 & \citet{Vrard18}\\
\noalign{\smallskip}                                                               
Distance (pc) & $645\pm 88$ & \citet{Wang16} &                                      &$\Delta\nu$ ($\mu$Hz) & $10.677\pm 0.016$ & \citet{Yu18}\\
Distance (pc) & $407^{+18}_{-18}$ & \citet{Huber17}\tablefootmark{b} &              &$\Delta\nu$ ($\mu$Hz) & $10.62\pm 0.05$ & \citet{Gaulme20}\\
\noalign{\smallskip}                                                                
Distance (pc) & $392.44\pm 12.70$ & \citet{Aguirre18} &            &$\nu_{\rm max}$ ($\mu$Hz) & $129\pm 4$ & \citet{Hekker11}\tablefootmark{a}\\
$T_{\rm{eff}}$ (K) & $4829^{+102}_{-102}$ & This work &                             &$\nu_{\rm max}$ ($\mu$Hz) & $127.52\pm 2.79$ & \citet{Pinsonneault14}\\
\noalign{\smallskip}                                                                
$T_{\rm{eff}}$ (K) & $4683\pm 77$ & \citet{Pinsonneault14} &                        &$\nu_{\rm max}$ ($\mu$Hz) & $128.607\pm 0.639$ & \citet{Huber17}\tablefootmark{b}\\
$T_{\rm{eff}}$ (K) & $4747\pm 5$ & \citet{Ness16} &                                 &$\nu_{\rm max}$ ($\mu$Hz) & $128.243\pm 0.012$ & \citet{Pinsonneault18}\\
$T_{\rm{eff}}$ (K) & $4747\pm 86$ & \citet{Huber17}\tablefootmark{b} &              &$\nu_{\rm max}$ ($\mu$Hz) & 130.0 & \citet{Vrard18}\\
$T_{\rm{eff}}$ (K) & $4770\pm 73$ & \citet{Pinsonneault18} &       &$\nu_{\rm max}$ ($\mu$Hz) & $128.74\pm 0.60$ & \citet{Yu18}\\
$T_{\rm{eff}}$ (K) & $4865\pm 100$ & \citet{Yu18} &                                 &$\nu_{\rm max}$ ($\mu$Hz) & $129.82\pm 0.25$ & \citet{Gaulme20}\\
$[\rm{Fe/H}]$ (dex) & $0.12^{+0.10}_{-0.11}$ & This work &                          &Age (Gyr) & $8.66\pm 1.06$ & \citet{Ness16}\\
$[\rm{Fe/H}]$ (dex) & $0.23\pm 0.05$ & \citet{Pinsonneault14} &                     &Age (Gyr) & $5.309\pm 0.001$ & \citet{Pinsonneault18}\\
$[\rm{Fe/H}]$ (dex) & $0.13\pm 0.11$ & \citet{Hawkins16} &                          &Age (Gyr) & $5.169\pm 1.270$ & \citet{Aguirre18}\\
$[\rm{Fe/H}]$ (dex) & $0.237\pm 0.004$ & \citet{Ness16} &                           &$M_{\ast}$ ($\rm{M_{\odot}}$) & $1.48\pm 0.20$ & \citet{Mosser14}\\
$[\rm{Fe/H}]$ (dex) & $0.157\pm 0.060$ & \citet{Huber17}\tablefootmark{b} &         &$M_{\ast}$ ($\rm{M_{\odot}}$) & $1.26^{+0.13}_{-0.11}$ & \citet{Pinsonneault14}\\
$[\rm{Fe/H}]$ (dex) & $0.224\pm 0.027$ & \citet{Pinsonneault18} &  &$M_{\ast}$ ($\rm{M_{\odot}}$) & $1.11\pm 1.02$ & \citet{Ness16}\\
$[\rm{Fe/H}]$ (dex) & $0.23\pm 0.15$ & \citet{Yu18} &                               &$M_{\ast}$ ($\rm{M_{\odot}}$) & $1.274\pm 0.046$ & \citet{Pinsonneault18}\\
$v_{\rm{turb}}$ ($\rm{km\,s^{-1}}$) & $1.13^{+0.27}_{-0.21}$ & This work &          &$M_{\ast}$ ($\rm{M_{\odot}}$) & $1.291\pm 0.061$ & \citet{Aguirre18}\\
\noalign{\smallskip}                                                                
$v_{\rm{turb}}$ ($\rm{km\,s^{-1}}$) & $1.037\pm 0.081$ & \citet{Hawkins16} &        &$M_{\ast}$ ($\rm{M_{\odot}}$) & $1.42\pm 0.041$ & \citet{Vrard18}\\
$v_{\rm{macro}}$ ($\rm{km\,s^{-1}}$) & $4.82^{+0.91}_{-4.82}$ & This work &         &$M_{\ast}$ ($\rm{M_{\odot}}$) & $1.38\pm 0.07$ & \citet{Yu18}\\
\noalign{\smallskip}                                                                
$v\,\sin i$ ($\rm{km\,s^{-1}}$) & $1.32^{+4.20}_{-0.53}$ & This work &              &$M_{\ast}$ ($\rm{M_{\odot}}$) & $1.42\pm 0.09$ & \citet{Gaulme20}\\
\noalign{\smallskip}                                                                
$\log g$ (dex) & $3.09^{+0.32}_{-0.33}$ & This work &                               &$M_{\ast}$ ($\rm{M_{\odot}}$) & $1.220^{+0.566}_{-0.565}$ & \citet{Sayeed21}\\
\noalign{\smallskip}                                                                
$\log g$ (dex) & $3.001\pm 0.011$ & \citet{Pinsonneault14} &                        &$R_{\ast}$ ($\rm{R_{\odot}}$) & $5.86^{+0.24}_{-0.22}$ & \citet{Pinsonneault14}\\
\noalign{\smallskip}                                                                
$\log g$ (dex) & $2.94\pm 0.01$ & \citet{Ness16} &                                  &$R_{\ast}$ ($\rm{R_{\odot}}$) & $5.854^{+0.104}_{-0.104}$ & \citet{Huber17}\tablefootmark{b}\\
\noalign{\smallskip}                                                                
$\log g$ (dex) & $3.015\pm 0.004$ & \citet{Huber17}\tablefootmark{b} &              &$R_{\ast}$ ($\rm{R_{\odot}}$) & $5.797\pm 0.016$ & \citet{Pinsonneault18}\\
$\log g$ (dex) & $3.017\pm 0.006$ & \citet{Pinsonneault18} &       &$R_{\ast}$ ($\rm{R_{\odot}}$) & $5.834\pm 0.121$ & \citet{Aguirre18}\\
$\log g$ (dex) & $3.016\pm 0.005$ & \citet{Aguirre18} &            &$R_{\ast}$ ($\rm{R_{\odot}}$) & $6.02\pm 0.11$ & \citet{Yu18}\\
$\log g$ (dex) & $3.021\pm 0.007$ & \citet{Yu18} &                                  &$R_{\ast}$ ($\rm{R_{\odot}}$) & $6.09\pm 0.12$ & \citet{Gaulme20}\\
$\log g$ (dex) & $3.02\pm 0.01$ & \citet{Gaulme20} &               &$R_{\ast}$ ($\rm{R_{\odot}}$) & $5.673^{+0.152}_{-0.146}$ & \citet{Sayeed21}\\
\noalign{\smallskip}                                                                
$\log g$ (dex) & $3.012\pm 0.200$ & \citet{Sayeed21} &             &Status & RGB\tablefootmark{c} & \citet{Mosser14}\\
$L$ ($\rm{L_{\odot}}$) & $16.4\pm 1.4$ & This work &                                &Status & RGB\tablefootmark{c} & \citet{Ness16}\\
$L$ ($\rm{L_{\odot}}$) & $15.716\pm 1.325$ & \citet{Aguirre18} &   &Status & RGB\tablefootmark{c} & \citet{Elsworth17}\\
 & & &                                                                              &Status & non He-core burning & \citet{Huber17}\tablefootmark{b}\\
 & & &                                                                              &Status & RGB\tablefootmark{c} & \citet{Vrard18}\\
 & & &                                                                              &Status & RGB\tablefootmark{c} & \citet{Yu18}\\
\hline
  \end{tabular}
\tablefoot{\tablefoottext{a}{OCT~I method with uncertainties based on synthetic results;}
\tablefoottext{b}{$\Delta\nu$ corrected direct method};
\tablefoottext{c}{RGB = red giant branch star.}}
\end{table*} 

The basic stellar parameters were determined from a high-resolution
(R=60,000) spectrum of KIC~3526061 taken without the iodine cell using 
the 2.7-m telescope at 
the McDonald Observatory with a S/N of $\sim 145$.
For the spectrum analysis we used the code \textsc{GSSP} \citep{Tkachenko15}
which works in a very fast and efficient way.
It employs the spectrum-synthesis method, by comparing the observed spectrum with a library 
of synthetic spectra computed by \textsc{SynthV} \citep{Tsymbal96} from atmosphere models 
on a grid of stellar parameters. \textsc{SynthV} is a spectrum synthesis code based on 
plane-parallel atmospheres and working in a non-local thermodynamic equilibrium regime. 
It has the advantage that for each chemical element different abundances can be considered.

The free stellar parameters in our analysis were the stellar effective
temperature, $T_{\rm{eff}}$, the stellar gravity, $\log g$, the
metallicity, $[\rm{Fe/H}]$, the microturbulent velocity, 
$v_{\rm{turb}}$, the macroturbulent velocity, $v_{\rm{macro}}$, and the
projected rotational velocity, $v\,\sin i$. The goodness of
the fit as well as the parameter uncertainties were calculated from
$\chi^2$-statistics \citep{Lehmann11}.

\textsc{GSSP} cannot adjust the observed continuum, except for a constant
factor, which has a limitation in analysing cool stars or the blue part of 
spectra that includes the higher Balmer lines. Therefore, we used it only in
a grid mode and used our own \textsc{MIDAS} programs for fitting \citep{Lehmann11}.
We restricted a wavelength range to 4369--5785 \AA~which provided best results
concerning a continuum value, S/N and overlap of the orders.

We encountered a large dependency between the projected rotational velocity
and the macroturbulent velocity. This is not surprising, because for slow rotators 
there is a trade off between the two parameters. 
Therefore we left both parameters free in the following analysis.
We determined $T_{\rm{eff}}$, $\log g$, $v_{\rm{turb}}$, $[\rm{Fe/H}]$, $v\,\sin
i$ and $v_{\rm{macro}}$ and then optimised the abundances of individual elements. Both steps 
were repeated in an iterative way. Resulting parameters are shown in 
Table~\ref{kic_stellar_par}, where we also list results of previous studies.
The resulting $v_{\rm{macro}}=4.82^{+0.91}_{-4.82}\,\rm{km\,s^{-1}}$ agrees 
very well with typical values for red giant stars of $\sim 5\,\rm{km\,s^{-1}}$
\citep{Gray88}. 

\begin{figure}
\centering
\includegraphics[width=\hsize]{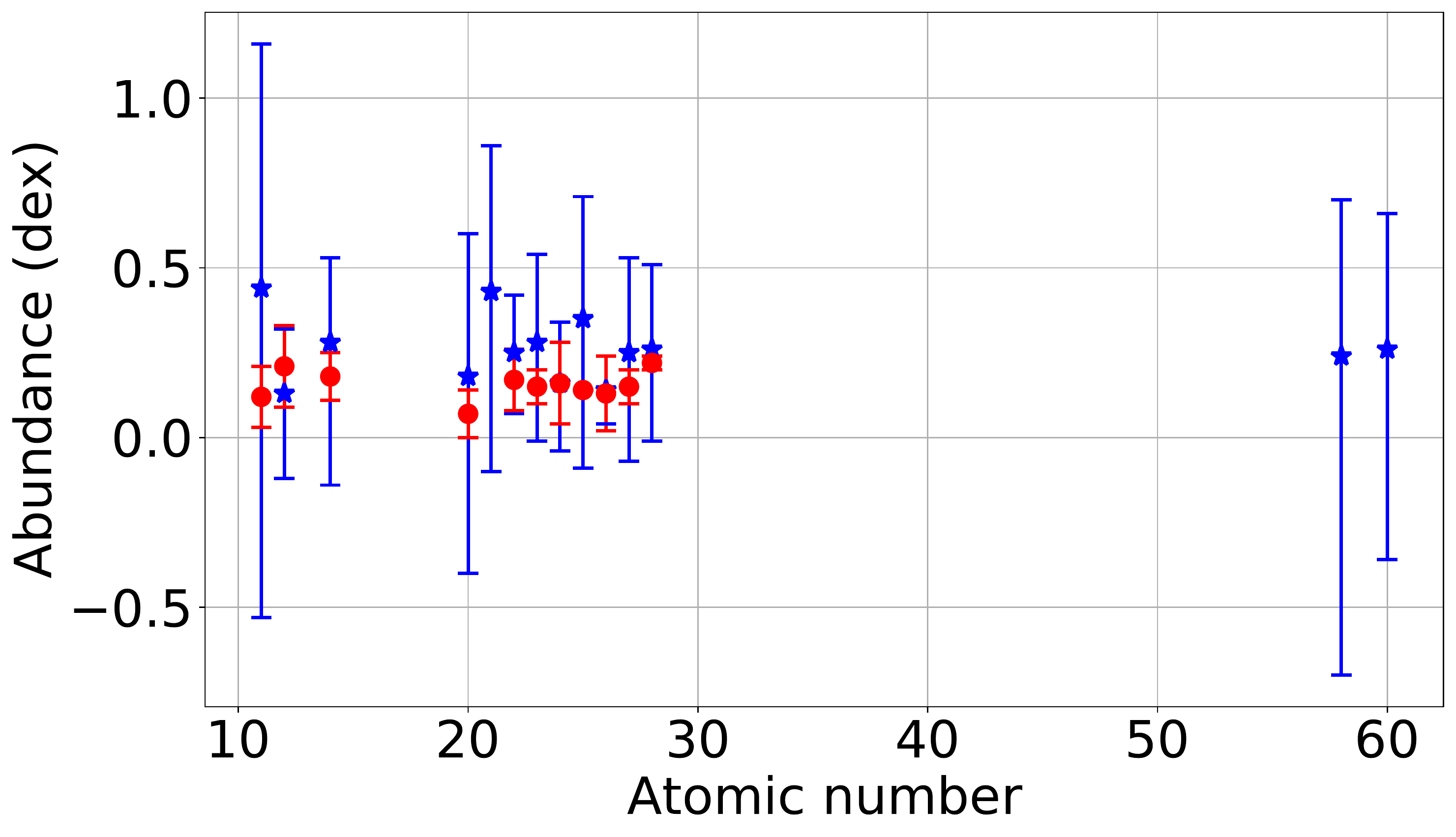}
\caption{Abundances of KIC~3526061 relative to solar composition
determined in this work (blue asterisks) and in \citet{Hawkins16} (red
filled circles).}
\label{Fig_abundances_kic}
\end{figure}

Results of abundances of chemical elements are listed in 
Table~\ref{kic_abundances}. All abundances determined in this work agree 
within error bars with the abundances from \citet{Hawkins16}, see also 
Fig.~\ref{Fig_abundances_kic}.

\begin{table*}\tabcolsep=7.4pt
\caption{Abundances of KIC~3526061 relative to solar composition 
determined in this work together with those known from literature.} \label{kic_abundances}
\centering
\begin{tabular}{lccccccc}
\hline 
\noalign{\smallskip}
Reference & Na & Mg & Si & Ca & Sc & Ti & V\\
\hline 
\noalign{\smallskip}
1 
& $+0.44^{+0.72}_{-0.97}$ 
& $+0.13^{+0.19}_{-0.25}$ 
& $+0.28^{+0.25}_{-0.42}$ 
& $+0.18^{+0.42}_{-0.58}$
& $+0.43^{+0.43}_{-0.53}$ 
& $+0.25^{+0.17}_{-0.18}$ 
& $+0.28^{+0.26}_{-0.29}$\\[1.5mm]
2 & $+0.12\pm 0.09$ & $+0.21\pm 0.12$ & $+0.18\pm 0.07$ & $+0.07\pm 0.07$ & &
$+0.17\pm 0.09$ & $+0.15\pm 0.05$\\
\hline
\noalign{\smallskip}
 & Cr & Mn & Fe & Co & Ni & Ce & Nd\\
\hline 
\noalign{\smallskip}
1 
& $+0.16^{+0.18}_{-0.20}$ 
& $+0.35^{+0.36}_{-0.44}$ 
& $+0.14^{+0.10}_{-0.10}$ 
& $+0.25^{+0.28}_{-0.32}$
& $+0.26^{+0.25}_{-0.27}$ 
& $+0.24^{+0.46}_{-0.94}$ 
& $+0.26^{+0.40}_{-0.62}$\\[1.5mm]
2 & $+0.16\pm 0.12$ & $+0.14\pm 0.00$ & $+0.13\pm 0.11$ & $+0.15\pm 0.05$ &
$+0.22\pm 0.02$ & & \\
\hline
  \end{tabular}
\tablebib{(1)~This work; (2)~\citet{Hawkins16}.}
\end{table*} 

\subsection{Brown dwarf candidate KIC~3526061~b}\label{orbit_kic}

\begin{figure}
\centering
\includegraphics[width=\hsize]{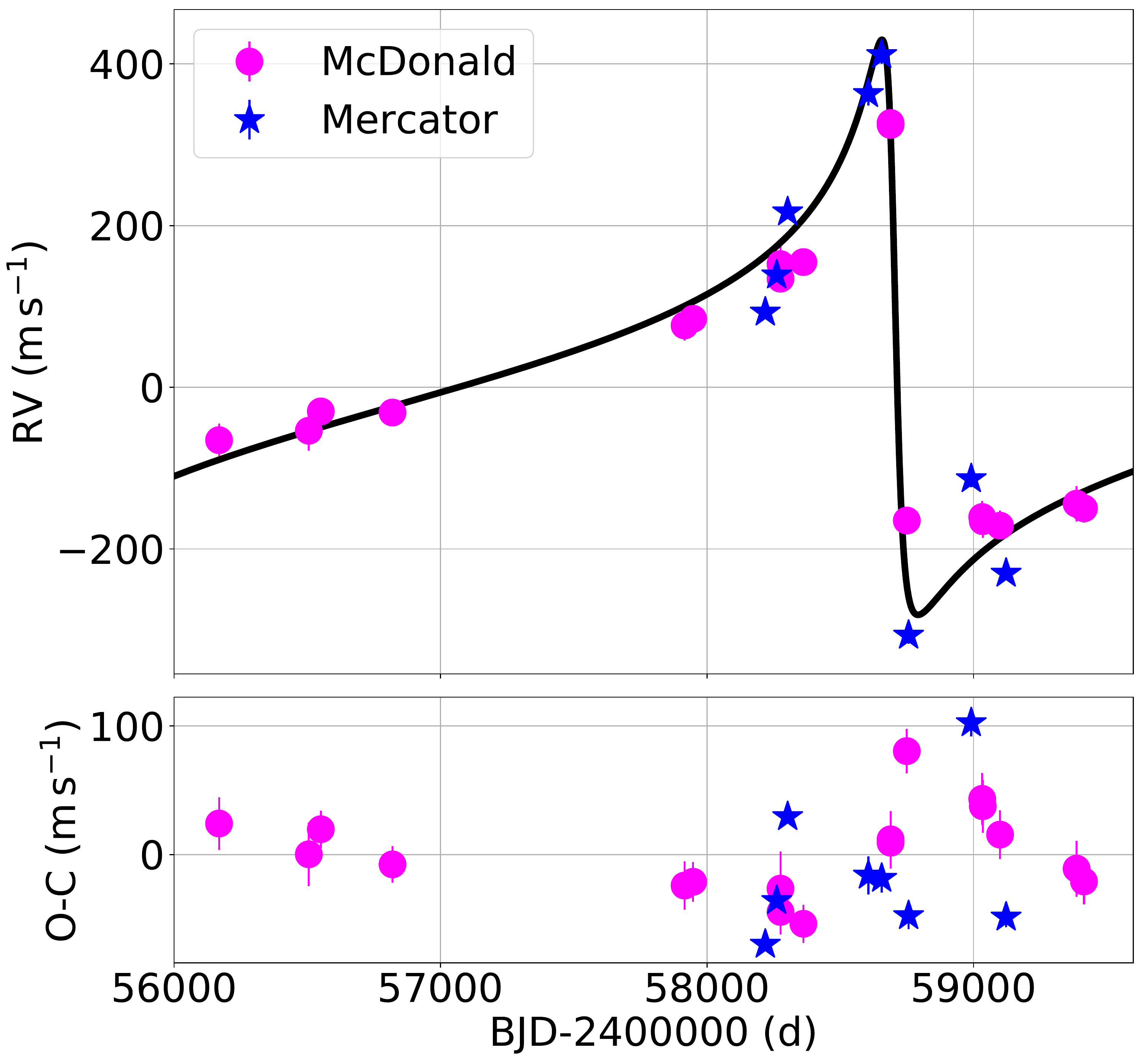}
\caption{RV measurements of KIC~3526061. Top: Data obtained from
August 2012 to July 2021 using the TS2 spectrograph at the McDonald
Observatory, Texas, and the HERMES spectrograph at Mercator, 
La Palma. The solid curve represents the Keplerian orbital solution. 
Bottom: RV residuals and error bars 
after removing the brown dwarf's orbital solution.}
\label{Fig_kic_orbital}
\end{figure}

\begin{figure}
\centering
\includegraphics[width=\hsize]{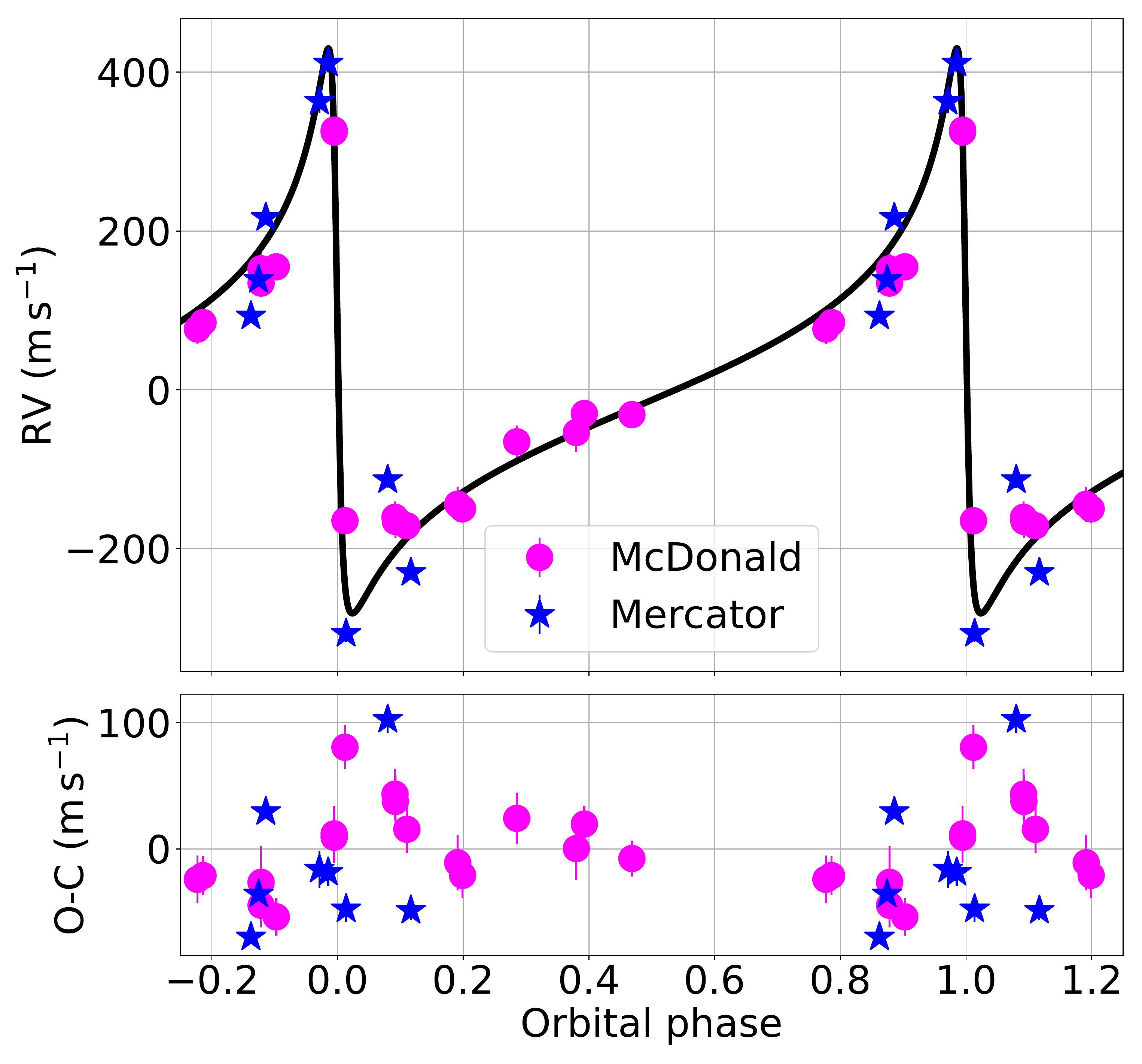}
\caption{Phased RV measurements of KIC~3526061. Top:
Data plotted with corresponding error
bars and phased to the orbital period of 3552 d. The Keplerian orbital solution is  
over-plotted with a solid curve. Bottom: RV residuals and error bars 
after removing the brown dwarf's orbital solution.}
\label{Fig_kic_phased}
\end{figure}

We monitored KIC~3526061 for nearly nine years and acquired 17 
RV measurements at the McDonald Observatory and eight RV measurements at the
Mercator telescope (see Table~\ref{kic_rvs} and Fig.~\ref{Fig_kic_orbital}). 
Our RV measurements show changes which could be caused by a sub-stellar companion. 
In order to access the nature of the companion we used the code
\textsc{pyaneti} \citep{Barragan19}, which employs a Bayesian approach
combined with Markov Chain Monte Carlo sampling to estimate the companion 
parameters. We derived the best-fitting orbital solution for KIC~3526061~b 
including the data from both data sets. A parameter space with 500 Markov chains was
explored to generate a posterior distribution of 20,000 independent points 
for each model parameter. 
We set uniform priors for all fitted parameters. We accounted for the RV
zero-points, $RV_0$, between the two different instruments and included jitter terms. 
We fitted for the orbital period,
$P$, time of minimum conjunction, $T_0$, eccentricity, $e$, periastron 
longitude, $\omega$, and semi-amplitude of the RV curve, $K$. The inferred
parameters are given in Table~\ref{kic_orbit}. They are defined as the
mean and 68 \% region of the credible interval of the posterior
distributions for each fitted parameter. 
In Table~\ref{kic_orbit}, the scatter of RV residuals 
around the orbital solution, rms, is given for all data and for 
data from each instrument.
We also used \textsc{FOTEL} \citep{Hadrava04} 
to check independently the orbital solution, and the resulting parameters were 
the same as for \textsc{pyaneti} within uncertainties. 
In Fig.~\ref{Fig_kic_orbital}, we show the RV measurements with the orbital
solution (top panel) and RV residuals with error bars (bottom panel).
Phase-folded RV variations for the orbital solution and the orbital 
fit are displayed in Fig.~\ref{Fig_kic_phased}.

\begin{figure}
\centering
\includegraphics[width=\hsize]{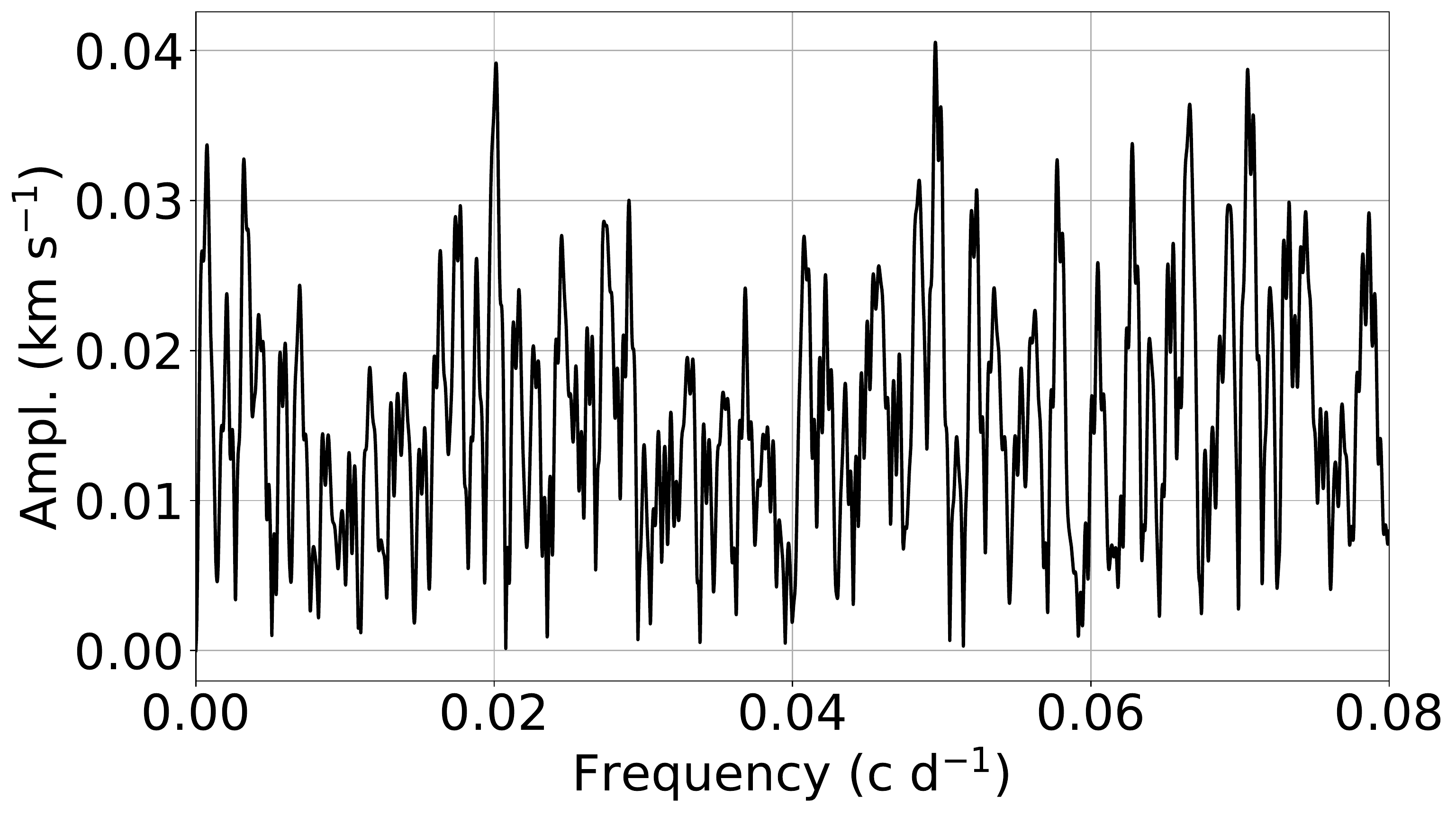}
\caption{Frequency amplitude spectrum of residual RV data of KIC~3526061 after
removing the Keplerian orbital solution.}
\label{Fig_kic_rv_resid}
\end{figure}

We searched for additional periods in residual RV data using the program
\textsc{Period04} \citep{Lenz05} based on the Fourier analysis, where multiple 
periods can be found via a pre-whitening procedure. A periodogram search out
to the Nyquist frequency in residual RV data 
showed no additional significant frequencies (see Fig.~\ref{Fig_kic_rv_resid}). 
The highest amplitude in the Fourier spectrum corresponds to a false alarm probability 
of greater than 50~\% using the criterion of \citet{Kuschnig97}.

\begin{table}
\caption{Orbital parameters of KIC~3526061~b.}\label{kic_orbit}
\centering
\begin{tabular}{lcc}\hline
Parameter & Value & Unit\\
\hline
\noalign{\smallskip}
\multicolumn{3}{l}{Fitted}\\
\noalign{\smallskip}
Period & $3552_{-135}^{+158}$ & d\\
\noalign{\smallskip}
$T_0$ & $2458708.1_{-1.9}^{+2.0}$ & d\\
\noalign{\smallskip}
$K$ & $355.7_{-5.0}^{+5.1}$ & $\rm{m\,s^{-1}}$\\
\noalign{\smallskip}
$e$ & $0.85_{-0.01}^{+0.01}$ & \\
\noalign{\smallskip}
$\omega$ & $75.7_{-1.3}^{+1.3}$ & deg\\
\noalign{\smallskip}
\multicolumn{3}{l}{Derived}\\
\noalign{\smallskip}
$M\,\sin i$ & $18.15_{-0.44}^{+0.44}$ & $\rm{M_{Jup}}$\\
\noalign{\smallskip}
$T_{\rm{periastron}}$ & $2458701.7_{-2.2}^{+2.4}$ & d\\
\noalign{\smallskip}
$a$ & $5.14\pm 0.16$ & AU \\
\noalign{\smallskip}
\multicolumn{3}{l}{Other parameters}\\
\noalign{\smallskip}
$RV_0^{\rm{McD}}$ & $9.7_{-5.9}^{+5.8}$ & $\rm{m\,s^{-1}}$\\
\noalign{\smallskip}
$RV_0^{\rm{HERMES}}$ & $-27557.9_{-3.3}^{+3.2}$ & $\rm{m\,s^{-1}}$\\
$\rm{rms\,_{Total}}$ & 40.6 & $\rm{m\,s^{-1}}$\\ 
$\rm{rms\,_{McD}}$ & 32.9 & $\rm{m\,s^{-1}}$\\
$\rm{rms\,_{HERMES}}$ & 53.3 & $\rm{m\,s^{-1}}$\\
\hline
\end{tabular}
\end{table} 

\subsubsection{Amplitude of stellar oscillations}

The scatter of RV residuals after fitting the companion's orbit is 
40.6~$\rm{m\,s^{-1}}$, which is higher than instrumental errors of our RV
measurements. This scatter cannot be solely due to solar-like oscillations. 
According to the scaling relations of \citet{Kjeldsen95}, the velocity amplitude 
for stellar oscillations is expected to be 
$v_{osc}=((L_{\ast}/{\rm{L_{\odot}}})/(M_{\ast}/{\rm{M_{\odot}}}))\,23.4\,\rm{cm\,s^{-1}}$. 
We derived the stellar luminosity 
$L_{\ast}=16.4\pm 1.4\,{\rm{L_{\odot}}}$ using our estimated stellar effective
temperature of $T_{\rm{eff}}=4829\,\rm{K}$ and the stellar radius of
\citet{Pinsonneault18}. We used this luminosity and the stellar mass 
of \citet{Vrard18} to calculate a velocity amplitude of
$v_{osc}=2.7\,\rm{m\,s^{-1}}$. We chose the stellar mass of \citet{Vrard18}
and the stellar radius of \citet{Pinsonneault18} for their smallest uncertainties 
of all values presented in Table~\ref{kic_stellar_par}.
We also used Eqs.~7 and 8 from \citet{Kjeldsen95} to predict a velocity amplitude
based on the luminosity amplitude from the \textit{Kepler} light curves, 
and derived $v_{osc}=1.9\,\rm{m\,s^{-1}}$.
Both velocity amplitude estimates are much smaller 
than the scatter of the RV residuals. The scatter of RV residuals 
is 32.9~$\rm{m\,s^{-1}}$ for the McDonald data and 53.3$\,\rm{m\,s^{-1}}$ for 
the Mercator data. As the McDonald data have lower scatter, we used them to
derive an orbital solution independent of the Mercator data. We followed a
similar procedure as above. The resulting best-fitting orbital period was
$P=4581\pm 400\,\rm{d}$ and eccentricity $e=0.91\pm 0.04$, and surprisingly
the scatter of RV residuals was only 9~$\rm{m\,s^{-1}}$. This value is
closer to our velocity amplitude estimates and is even lower than the
instrumental errors of RV measurements. This indicates good long-term stability
of RVs from the TS2 spectrograph at the McDonald Observatory over the time period of
nine years. 

However, the period based on the 
McDonald data only is quite different than the one obtained on the combined data. 
This shows that in fact an uncertainty of the period is much larger than the
one given in Table~\ref{kic_orbit}. The large error of the orbital period arises
from a combination of having a time base of observations close to the orbital period and 
sparse sampling of observations around the periastron passage. 
It is obvious that to refine the orbital solution more RV measurements from stable 
spectrographs such as TS2 at the McDonald Observatory would be needed, 
and particularly observations around the time of the predicted periastron passage.

\subsubsection{Mass of a companion}

Using a stellar mass of $1.42\pm 0.041\,\rm{M_{\odot}}$ \citep{Vrard18}
the derived minimum mass of the companion is 
$M\,\sin i=18.15\pm 0.44\,\rm{M_{Jupiter}}$. 
\citet{Jorissen20} classified KIC~3526061 as a
spectroscopic binary based on unpublished RVs. We have used
their data to make a common orbital solution with our data, which resulted to a 
scatter of RV residuals after fitting the companion's orbit
of 53.4~$\rm{m\,s^{-1}}$. This is a larger scatter
than without including the data of \citet{Jorissen20}. 
Also, their RVs have uncertainties of $\sim
45\,\rm{m\,s^{-1}}$, which is larger than for our data. For these two
reasons we have not included their data into our analysis. However, 
we confirm that their data are consistent with our results, giving the 
derived minimum mass of the companion in the common orbital solution of 
$18.27\pm 0.44\,\rm{M_{Jupiter}}$.

The unknown orbital
inclination leaves an open question as to the real nature of the companion.
Under the assumption that inclination angles are randomly distributed 
on the sky, there is only 2.6 \% chance for the companion's mass to be
greater than $80\,\rm{M_{Jupiter}}$, which has been proposed as a dividing
line between sub-stellar and stellar objects \citep{Burrows01, Hatzes15b, Chen17}.
This means that very likely KIC~3526061~b is a sub-stellar object, either a
brown dwarf or a giant planet. The masses of brown dwarfs have been defined 
as being in the range 13 -- 80~$\rm{M_{Jupiter}}$ \citep{Burrows01}, where
objects sustain deuterium burning through nuclear fusion for typically 0.1
million years, but are below the ignition limit of hydrogen at 75 --
80~$\rm{M_{Jupiter}}$. Another division between giant planets and brown
dwarfs is based on a formation, where the mass domains overlap 
since the minimum brown-dwarf mass is a few Jupiter masses and the maximum 
exoplanet mass can be as high as $\sim 30\,\rm{M_{Jupiter}}$
\citep{Chabrier14}.
\citet{Whitworth18} argued that, as regards their formation, brown
dwarfs should not be distinguished from hydrogen-burning stars. 
\citet{Hatzes15b}, on the other hand, suggested that, based on the mass-density 
relationship of sub-stellar objects, all objects in the mass range 
0.3 -- 60 $\rm{M_{Jupiter}}$ should be considered as giant planets.
This was further corroborated by
\citet{Chen17} who found that brown dwarfs follow the same trend as giant
planets in the mass-radius diagram. 

\subsubsection{Transit probability, depth and duration}

In order to calculate the probability, depth and duration of a potential
transit, we first estimated the sub-stellar companion's radius. We used the 
mass-radius relationship for the Jovian regime in a form $R_P \propto\,M_P^{-0.04}$
given by \citet{Chen17}, which resulted in the companion's radius 0.89
$\rm{R_{Jupiter}}$. Assuming an inclination of $i=90^{\circ}$ and ignoring 
limb darkening effects we calculated a transit probability of 3.5 \%, using
Eq.~5 from \citet{Kane08}, a transit depth of 0.0244 \% or 244 ppm, and a
transit duration of 0.9 d. KIC~3526061 as an evolved star presents
a photometric variability of about 700 ppm in the \textit{Kepler} light
curve, and that is larger than the expected transit depth 
and complicates searching for a potential transit signal. 
The real difficulty, however, comes from the closeness of the transit 
duration and a time scale of intrinsic variations making them difficult to
disentangle.
Moreover, currently we do not have a good
knowledge about potential transit times due to the large uncertainty in the orbital
period. 

\subsubsection{Wide orbit}

\begin{figure}
\centering
\includegraphics[width=\hsize]{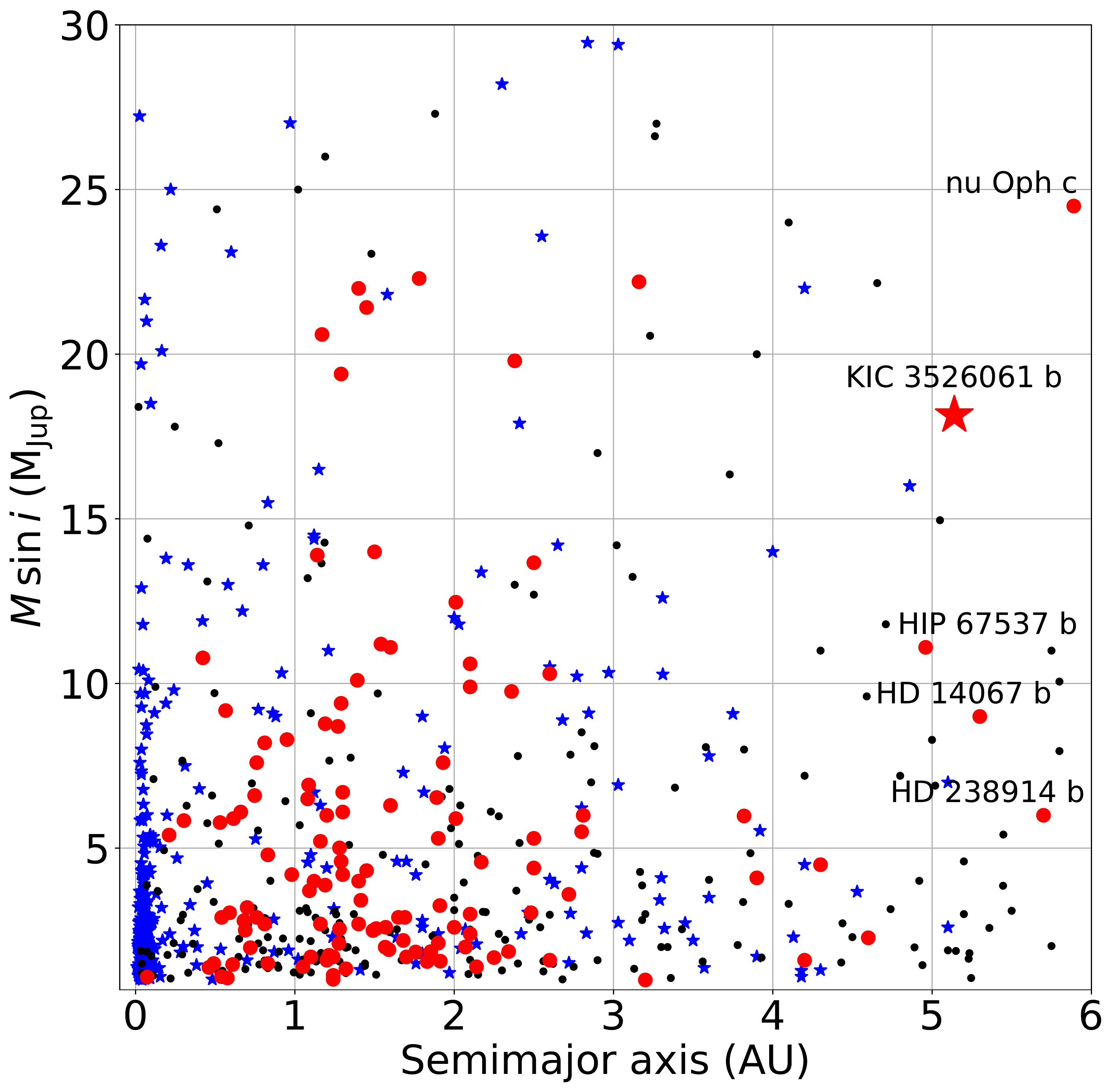}
\caption{Minimum planet mass versus semi-major axis for known giant planets
($M\,\sin i\geq 1.0\,\rm{M_{Jupiter}}$) with a semi-major
axis smaller than 6~AU.
The black dots and the blue asterisks correspond to main sequence host 
stars (http://exoplanet.eu/) and the red filled circles correspond to giant host 
stars (https://www.lsw.uni-heidelberg.de/users/sreffert/giantplanets/giantplanets.php).
The blue asterisks are transiting systems, for which the true
mass of the companion is shown. The red asterisk shows the position of KIC~3526061~b.}
\label{Fig_msinivsa}
\end{figure}

KIC~3526061~b has one of the longest orbital period of sub-stellar
companions that orbit a giant star known so far. We show the position of KIC~3526061~b
in the semi-major axis versus minimum planet mass diagram in Fig.~\ref{Fig_msinivsa}. 
KIC~3526061~b is placed in a barely populated region of the diagram. 
More or similarly distant companions were discovered by 
\citet{Quirrenbach11} around $\nu$~Oph with a companion's minimum mass of 
$24.5\,\rm{M_{Jupiter}}$ at the orbital distance of a~=~5.89~AU, 
\citet{Wang14} around HD~14067 with a companion's minimum mass of $9\,\rm{M_{Jupiter}}$ 
at the orbital distance of a~=~5.3~AU, \citet{Jones17} around HIP~67537 with 
a companion's minimum mass of $11.1\,\rm{M_{Jupiter}}$ at the orbital distance 
of a~=~4.91~AU, and \citet{Adamow18} around HD~238914 with a companion's
minimum mass of $6\,\rm{M_{Jupiter}}$ at the orbital distance of a~=~5.7~AU.

\subsubsection{Origin of a large orbital eccentricity}

\begin{figure}
\centering
\includegraphics[width=\hsize]{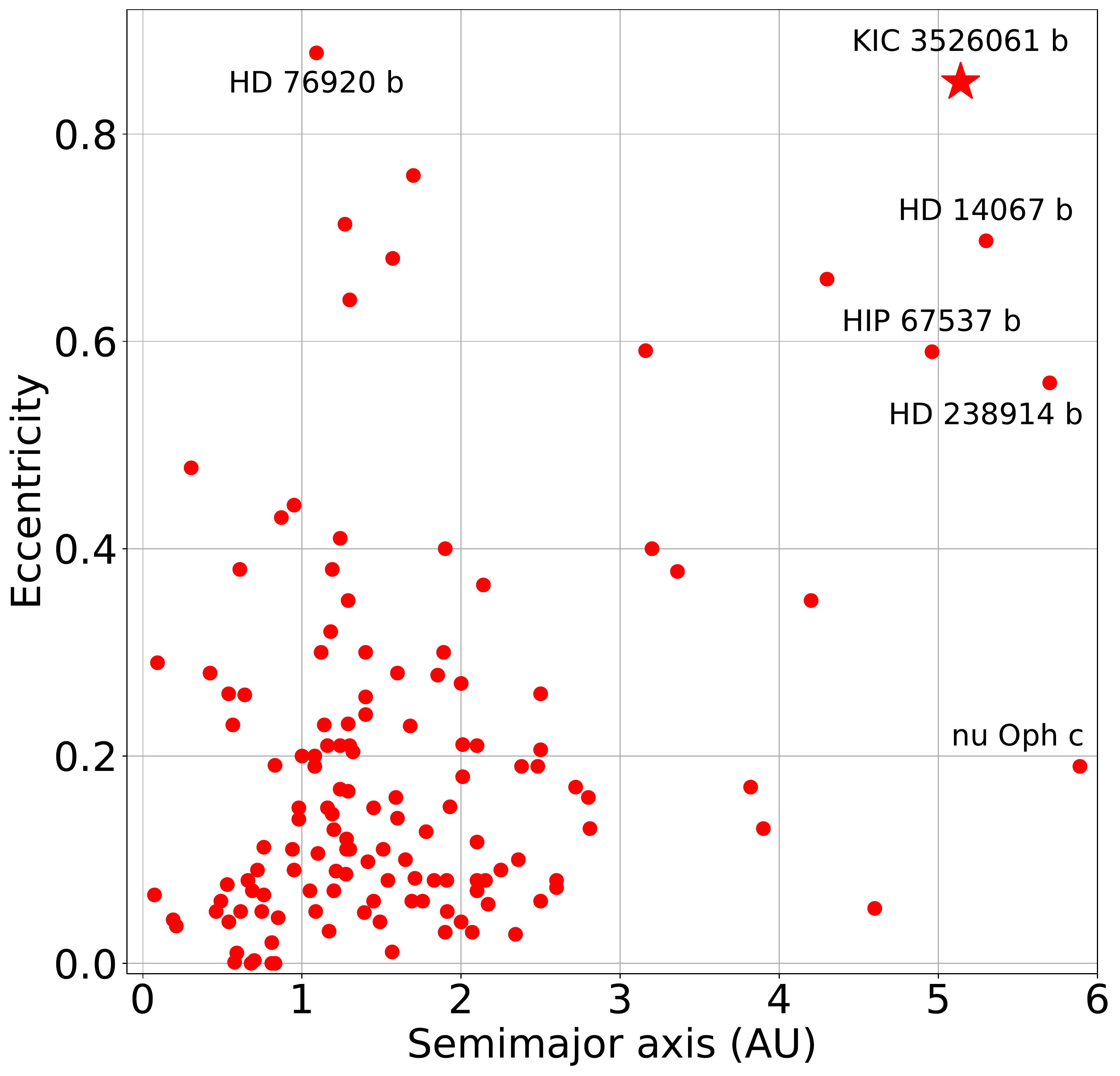}
\caption{Eccentricity versus semi-major axis for all sub-stellar companions known around
giant host stars 
(source: https://www.lsw.uni-heidelberg.de/users/sreffert/giantplanets/giantplanets.php).
The red asterisk shows the position of KIC~3526061~b.}
\label{Fig_evsa}
\end{figure}

KIC~3526061 is the most evolved system found having a sub-stellar companion with such a
large eccentricity and wide separation (see Fig.~\ref{Fig_evsa}).
The orbital eccentricity $0.85\pm 0.01$ of KIC~3526061~b is the second
largest of sub-stellar companions orbiting giant stars.
So far only one planetary companion HD~76920~b with 
$M\,\sin i=3.13_{-0.43}^{+0.41}\,\rm{M_{Jupiter}}$ and semimajor axis 1.091
AU orbiting a red giant star was found to have larger eccentricity of 
$0.8782\pm 0.0025$ \citep{Wittenmyer17, Bergmann21}. 

The origin of such a highly eccentric orbit is linked with the formation and
dynamical history of KIC~3526061~b. The following two main mechanisms for the 
formation of wide-orbit giant
companions within protoplanetary disks have been proposed: 
top-down formation by gravitational disk instability \citep[e.g.][]{Boss97}, 
and bottom-up formation by core accretion \citep[e.g.][]{Pollack96}.
Planets are expected to form on circular coplanar orbits within
protoplanetary disks via a core accretion, but can develop nonzero
eccentricities through a planet-planet scattering \citep[e.g.][]{Chatterjee08,Ford08}, 
secular Kozai-Lidov perturbations 
with a massive outer companion \citep[e.g.][]{Naoz16}, or planet-disk
interactions \citep[e.g.][]{Goldreich03}. However, there are no indications 
for additional massive companions in the RV data of KIC~3526061.
We also used the SIMBAD astronomical database \citep{Wenger00}
and checked all sources within 5 arcmins from KIC~3526061, and found no
targets with compatible parallax and proper motions. The scenario of a
captured free-floating sub-stellar companion is also unlikely. According to 
simulations by \citet{Parker17}, free-floating planets are usually 
captured into much wider orbits. In a planet-planet scattering scenario, 
a second sub-stellar companion of a comparable mass would have
either been ejected from the system as the result of a close encounter with
KIC~3526061~b, or pushed outwards into a long-period orbit that is beyond
our current detection limit, or engulfed by the star. The parameter space of
a putative second companion is limited by the highly eccentric orbit of
KIC~3526061~b, which cruises between 0.77 -- 9.51 AU, or 28 -- 349~$R_{\ast}$.

KIC~3526061~b is a sub-stellar companion on the wide orbit, where occurrence
rates are low and there are orders of magnitude fewer discoveries compared
to short-period systems, therefore population-level studies might give us a better 
understanding of the formation of KIC~3526061~b.     
\citet{Bowler20} combined new high-contrast imaging observations with astrometry 
to test for differences in the population-level eccentricity
distributions of 27 long-period giant planets and brown dwarfs between 5 and
100 AU. Their analysis revealed that low-mass companions ($<15\,\rm{M_{Jupiter}}$)
and low-mass ratio systems ($M_2/M_1<0.01$) preferentially have lower eccentricities,
similar to the population of warm Jupiters at small separations, and
the brown dwarf companions (15 -- 75 $\rm{M_{Jupiter}}$) and higher-mass
ratio systems ($M_2/M_1=0.01$ -- 0.2) exhibit higher eccentricities.
Their explanation is that these populations predominantly form in
distinct manners: the planetary-mass companions originate in
disks and form via a core accretion, while brown dwarf companions represent the
low-mass ratio end of binary star formation.

The mass ratio of the KIC~3526061 system is 0.0122, which places it to
higher-mass ratio systems according to \citet{Bowler20} and points to the
formation of KIC~3526061~b via a gravitational instability similarly as 
for the star formation. This is supported also by other studies.  
\citet{Nielsen19} presented statistical results from the ﬁrst 300 stars in the
GPIES survey. They found that giant planets follow a bottom-heavy mass distribution 
and favour smaller semimajor axes, while brown dwarfs exhibit just the opposite 
behaviour, which points to formation of giant planets by core/pebble accretion, 
and formation of brown dwarfs by gravitational instability.
\citet{Wagner19} analysed the underlying relative mass distribution of sub-stellar 
companions using survival analysis and concluded that core accretion 
is the primary mechanism at forming companions less massive than 
$\sim$~10 -- 20~$\rm{M_{Jupiter}}$, and that gravitational instability
is the primary mechanism at forming higher-mass companions.
\citet{Ma14} found that brown dwarfs with masses
lower than $\sim 43\,\rm{M_{Jupiter}}$ have an eccentricity distribution 
consistent with giant planets in the mass-eccentricity diagram, while brown
dwarfs with masses above $\sim 43\,\rm{M_{Jupiter}}$ have the star-like 
eccentricity distribution. They concluded that these results support the
idea that
brown dwarfs below this mass limit form in protoplanetary discs
around host stars and above this mass limit form like stellar binary
systems. They also noted that their sample is not sufficient to exclude a
possibility that a small number of brown dwarfs in each of the two mass regions 
may form in an opposite formation mechanism. 

Based on the population-level studies it seems more likely that KIC~3526061~b was
formed via a gravitational instability, which is also consistent with 
findings of \citet{Ma14}, because we have only a minimum mass estimate 
for KIC~3526061~b. However, it is not excluded that KIC~3526061~b formed
via a core accretion and developed such a large eccentricity via e.g. a planet-planet
scattering. In addition, most direct imaging surveys have preferentially focused on young
stars and it is not clear whether the population-level eccentricities 
of sub-stellar companions evolve over time, or they are established at young ages
\citep{Bowler20}. KIC~3526061 as the most evolved system found having a sub-stellar 
companion with such a large eccentricity and wide separation (see Fig.~\ref{Fig_evsa}) 
might provide a probe of the dynamical evolution of such systems over time. 

\subsection{Stellar activity analysis of KIC~3526061}\label{activity_kic}

\subsubsection{Photometric variations}\label{activityphot_kic}

We analysed the \textit{Kepler} photometry of KIC~3526061 in order to check
whether there are any stellar activity features such as rotational modulation  
seen in the light curve. The star was observed by the \textit{Kepler} 
satellite \citep{Borucki10} since May 2009 and for all quarters during the 
main \textit{Kepler} mission. We used presearch data conditioning simple 
aperture photometry (PDCSAP) light curves \citep{Smith12,Stumpe12} downloaded from 
the Mikulski Archive for Space Telescopes \citep{Thompson16}. 
In order to remove systematic trends in light curves, multi-scale cotrending 
basis vectors were used. They were built from the most common systematic trends 
observed in each quarter of the \textit{Kepler} data. For light
curves retrieval, we made use of the following packages: \textsc{lightkurve}
\citep{Cardoso18}, \textsc{astropy} \citep{Price-Whelan18} and
\textsc{astroquery} \citep{Ginsburg19}. 
The orbital period of KIC~3526061~b is
approximately twice as long as the time span of the \textit{Kepler} data.
We have not detected any trends signalising photometric variations with
the orbital period.

KIC~3526061 has also been observed in sectors 14 and 26 of the Transiting 
Exoplanet Survey Satellite \citep[TESS;][]{Ricker14}. We used PDCSAP light curves 
as above for the \textit{Kepler} data, and found neither long-term
variations nor evidence of transit events. 

\subsubsection{Spectral line bisector analysis}\label{activitybis_kic}

\begin{figure}[thb]
\centering
\includegraphics[width=\hsize]{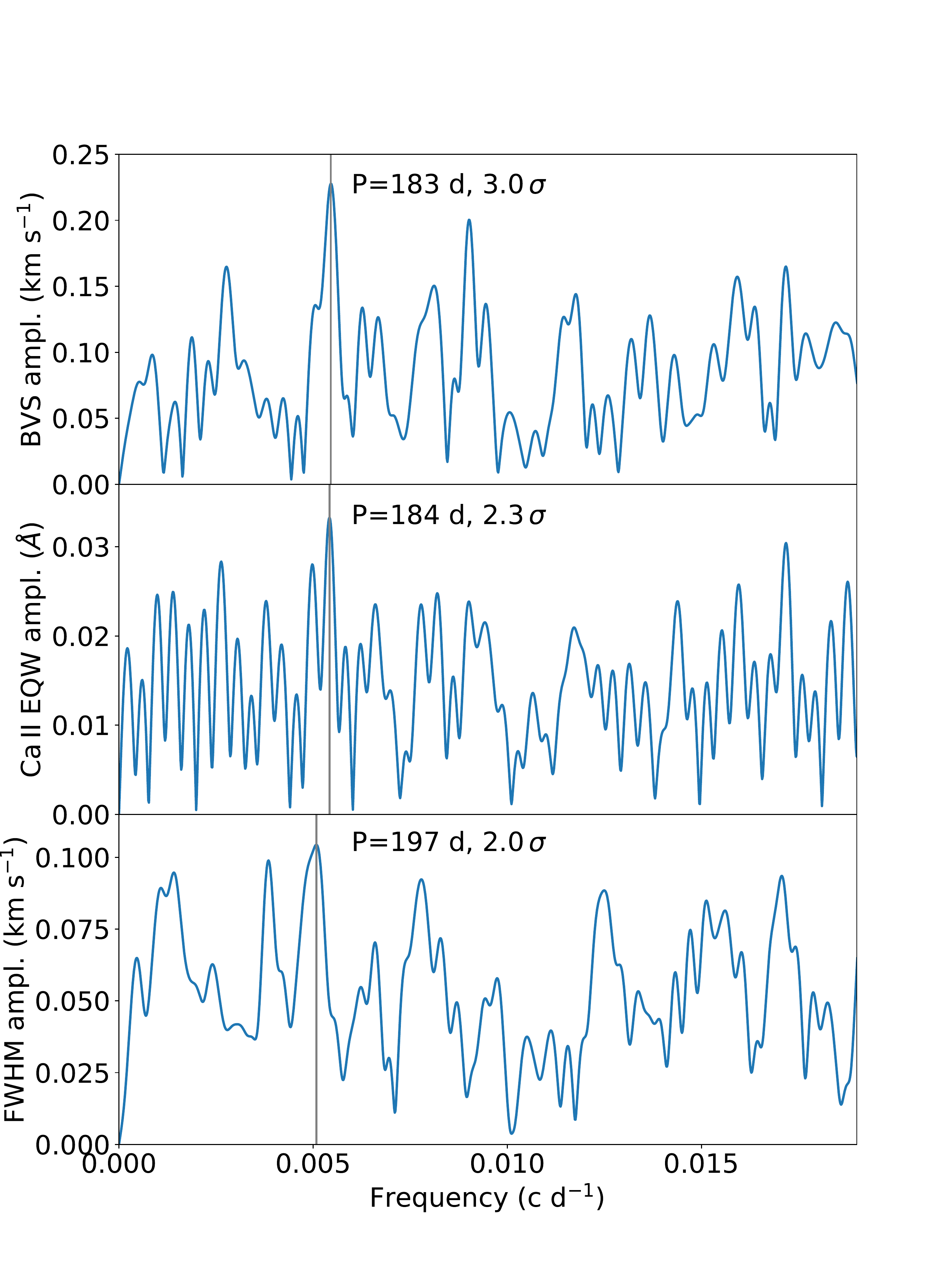}
\caption{Frequency amplitude spectra for KIC~3526061. 
Top: Bisector velocity spans. Middle: Equivalent width of the \ion{Ca}{ii} 
triplet lines. Bottom: Full-width at half maximum of spectral line shapes. 
For all three quantities we found variations on a similar time scale.}
\label{Fig_BVS_FWHM_EQW}
\end{figure}

Inhomogeneous features on the stellar surface can create variable asymmetries 
in spectral line profiles as a star rotates. This asymmetry can be described 
with the line bisector, which consists of midpoints of horizontal line segments 
extending across a line profile. Typically, due to stellar granulation, the red 
wing of a stellar line is depressed, which causes a bisector to have a positive 
slope and curve to the right near the continuum level.

For the McDonald data, we used fxcor task in IRAF to derive a 
cross-correlation function (CCF) for each
spectrum. We used a spectral range 4440 -- 4660, 6100 -- 6260 and 6340 --
6427~\AA, where no iodine and telluric lines were present, and where we obtained the most 
accurate results. Then we calculated a bisector for each individual CCF.

For the Mercator data, we used a CCF of the \textsc{HermesDRS} data
reduction pipeline to calculate bisectors. The CCF was typically based on 
$\sim$1655 spectral lines for each spectrum which results in a very good 
accuracy for the bisector measurement. 
As we are interested in relative and not absolute bisector measurements,
using an average of many lines is appropriate for studying variations 
of line bisectors with time \citep[e.g.][]{Martinez05, Nowak13}.
We used normalised bisectors for both data sets for the subsequent analysis. 

Bisector changes can be detected from bisector velocity span (BVS)
measurements, which consist of measuring the difference between bisectors 
at two different flux levels of a spectral line \citep{Hatzes15a}.
We measured the BVS of the spectral profile using the difference of the
average bisector values between flux levels of 0.16 -- 0.36 and 0.68 -- 0.88
of the continuum value, which means that we avoided the spectral core and wing, 
where errors of the bisector measurements are larger. 

We searched for periods in the BVS variations using the program
\textsc{Period04} \citep{Lenz05}. The frequency spectrum is displayed in the top 
panel of Fig.~\ref{Fig_BVS_FWHM_EQW}. The most significant peak was found at the
frequency $0.005453\,\rm{c\,d^{-1}}$, corresponding to a period of
$P=183\,\rm{d}$, with a 3.0 $\sigma$ significance and a 
false alarm probability of 0.016.

\subsubsection{Spectral line shape analysis}\label{activityfwhm_kic}

In addition to the spectral line bisector analysis, there are other
quantities which can be measured to evaluate the stellar activity over the
rotation period of the star. We measured a full-width at half maximum (FWHM)
of each CCF derived as described in the previous section
to access if there are any changes of spectral line shapes.    

We searched for periods in the FWHM variations using the program
\textsc{Period04} \citep{Lenz05}. The frequency spectrum is displayed in the
bottom panel of Fig.~\ref{Fig_BVS_FWHM_EQW}. The most significant peak was found at the
frequency $0.00508\,\rm{c\,d^{-1}}$, corresponding to a period of
$P=197\,\rm{d}$, with a 2.0 $\sigma$ significance.

\subsubsection{Chromospheric activity}\label{activitycaII_kic}

The equivalent width (EQW) variations of the \ion{Ca}{ii} H \& K lines are often
used as a chromospheric activity indicator. They are sensitive to stellar
activity which in turn may affect the measured RV variations. In
chromospherically active stars, the \ion{Ca}{ii} H \& K lines show a typical line
core reversal \citep{Pasquini88}. Spectra of KIC~3526061 from both sites
have a low S/N in a region of the \ion{Ca}{ii} H \& K lines,
which complicates an analysis of possible emission features in line cores. 
Instead, the \ion{Ca}{ii} triplet lines are often used to measure chromospheric 
activity \citep{Hatzes03, Lee13}.

\citet{Linsky79} showed that \ion{Ca}{ii} 8542 \AA~line was suitable as a 
diagnostic of stellar chromospheric activity. The \ion{Ca}{ii} 8662 \AA~line
is also suitable for this purpose and, unlike \ion{Ca}{ii}
8498 \AA~and \ion{Ca}{ii} 8542 \AA, is uncontaminated by atmospheric water
vapour lines near the line core \citep{Larson93}.

For the McDonald data, we measured the EQW at the central part of the
\ion{Ca}{ii} 8542 \AA~line from 8540.97 to 8543.24 \AA~and of the \ion{Ca}{ii} 8662 \AA~line 
from 8661.0 to 8663.27 \AA. We averaged the two measurements, and searched for periods
using the program \textsc{Period04} \citep{Lenz05}. The frequency spectrum is displayed in the
middle panel of Fig.~\ref{Fig_BVS_FWHM_EQW}. The most significant peak was found at the
frequency $0.005422\,\rm{c\,d^{-1}}$, corresponding to a period of
$P=184\,\rm{d}$, with a 2.3 $\sigma$ significance. 

In the Mercator data, we measured the EQW at the central part of the 
\ion{Ca}{ii} 8498 \AA~line from 8497.52 to 8498.58 \AA, of the
\ion{Ca}{ii} 8542 \AA~line from 8541.63 to 8542.60 \AA~and of the 
\ion{Ca}{ii} 8662 \AA~line in two different ranges, 8661.55 to 8662.73 \AA~and 
8661.0 to 8663.27 \AA. We did not find any significant peaks in frequency
spectra. This could be also due to a small number of measurements; only
eight spectra were taken of KIC~3526061 at the Mercator telescope.

\subsubsection{Rotation period of KIC~3526061}\label{rotper_kic}

In Fig.~\ref{Fig_BVS_FWHM_EQW} we show frequency amplitude spectra of our measurements
of bisector velocity spans, a full-width at half maximum of spectral line
shapes and equivalent width of the \ion{Ca}{ii} triplet lines (for details see 
previous sections). We found variations on a similar time scale of about 183
days in all three quantities. If these variations would be related to a stellar 
rotation period, and we assume a stellar rotation period of 183 days and 
the stellar radius of \citet{Pinsonneault18}, 
then the projected rotational velocity would be $v\,\sin i=1.6\,\rm{km\,s^{-1}}$. This
would lift the degeneracy described in Sect.~\ref{properties_kic} between the 
projected rotational velocity and the macroturbulent velocity.
However, we should be cautious because the signal is weak, is based only on
25 observations spread over a long time of nine years, and in addition is
about half of the year period. In fact the variations can be due to 
changes of the instrumental profile which varies between individual observing runs 
and is the most likely cause of variations in all three measured quantities. 
This is specially true of the McDonald data since there are moving
components in the Tull Spectrograph such as the echelle grating and prisms
which are changed for different setups according to the observer's needs. 
At this moment, we cannot resolve on what is the real reason of the
variations. 

\section{HD~187878}\label{HD187878}

\subsection{Stellar properties}\label{properties_hd}

HD~187878 has a visual magnitude of $m_V=7.13\pm 0.01$ mag \citep{Hog00}. 
The parallax was determined from \textit{Gaia} EDR3 data
as $5.1501\pm 0.0597$ mas \citep{Gaia16, Gaia21a}, which implies an absolute magnitude 
$M_V=0.69\pm 0.03$ mag. Table~\ref{hd_stellar_par} lists stellar parameters 
known from literature together with those determined in this work.

\begin{table}\tabcolsep=3.0pt
\caption{Stellar parameters of HD~187878 from this work together with
those known from literature.}\label{hd_stellar_par}
\centering
\begin{tabular}{lll}
\hline 
\noalign{\smallskip}
Parameter & Value & Reference\\
\noalign{\smallskip}
\hline 
\noalign{\smallskip}
$m_V$ (mag) & $7.13\pm 0.01$ & \citet{Hog00}\\
$B-V$ (mag) & $0.98\pm 0.02$ & \citet{Hog00}\\
Parallax (mas) & $5.1501\pm 0.0597$ & Gaia Coll. et al. (2021a)\\
$M_V$ (mag) & $0.69\pm 0.03$ & This work\\
Distance (pc) & $194\pm 2$ & This work\\
Distance (pc) & 201.75887 & \citet{Cruzalebes19}\\
$T_{\rm{eff}}$ (K) & $5168^{+65}_{-63}$ & This work\\
$T_{\rm{eff}}$ (K) & 5053 & \citet{McDonald12}\\
$T_{\rm{eff}}$ (K) & $5091\pm 80$ & \citet{Thygesen12}\\ 
$T_{\rm{eff}}$ (K) & 5003.38 & \citet{Cruzalebes19}\\
$[\rm{Fe/H}]$ (dex) & $0.00^{+0.06}_{-0.07}$ & This work\\
\noalign{\smallskip}
$[\rm{Fe/H}]$ (dex) & $0.00\pm 0.15$ & \citet{Thygesen12}\\ 
$v_{\rm{turb}}$ ($\rm{km\,s^{-1}}$) & $1.56^{+0.12}_{-0.16}$ & This work\\
$v_{\rm{turb}}$ ($\rm{km\,s^{-1}}$) & $1.37\pm 0.15$ & \citet{Thygesen12}\\
$v_{\rm{macro}}$ ($\rm{km\,s^{-1}}$) & $0.6 \pm 1$ & \citet{Thygesen12}\\
$v\,\sin i$ ($\rm{km\,s^{-1}}$) & $4.79^{+0.46}_{-0.46}$ & This work\\
$v\,\sin i$ ($\rm{km\,s^{-1}}$) & $4.5 \pm 1$ & \citet{Thygesen12}\\
$\log g$ (dex) & $2.89^{+0.19}_{-0.19}$ & This work\tablefootmark{a}\\ 
$\log g$ (dex) & $2.776 \pm 0.004$ & This work\tablefootmark{b}\\
$\log g$ (dex) & $2.80\pm 0.01$ & \citet{Thygesen12}\\
$\log g$ (dex) & $2.746\pm 0.067$ & \citet{Ghezzi15}\tablefootmark{c}\\
$\log g$ (dex) & $2.800\pm 0.022$ & \citet{Ghezzi15}\tablefootmark{b}\\
$\log g$ (dex) & $2.77\pm 0.01$ & \citet{Gaulme20}\\
$L$ ($\rm{L_{\odot}}$) & $80.2\pm 9.5$ & This work\\
$L$ ($\rm{L_{\odot}}$) & 66.67 & \citet{McDonald12}\\ 
$\Delta\nu$ ($\rm{\mu\,Hz}$) & $6.09\pm 0.07$ & This work\\
$\Delta\nu$ ($\rm{\mu\,Hz}$) & $6.12\pm 0.12$ & \citet{Hekker11}\\ 
$\Delta\nu$ ($\rm{\mu\,Hz}$) & $6.18\pm 0.05$ & \citet{Gaulme20}\\
$\nu_{\rm max}$ ($\rm{\mu\,Hz}$) & $70.4\pm 0.4$ & This work\\
$\nu_{\rm max}$ ($\rm{\mu\,Hz}$) & $76.00\pm 3.80$ & \citet{Hekker11}\\
$\nu_{\rm max}$ ($\rm{\mu\,Hz}$) & $70.70\pm 0.45$ & \citet{Gaulme20}\\
$M_{\ast}$ ($\rm{M_{\odot}}$) & $2.6 \pm 0.1$ & This work\\
$M_{\ast}$ ($\rm{M_{\odot}}$) & $2.162\pm 0.236$ & \citet{Ghezzi15}\tablefootmark{c}\\
$M_{\ast}$ ($\rm{M_{\odot}}$) & $2.950\pm 0.504$ & \citet{Ghezzi15}\tablefootmark{b}\\
$M_{\ast}$ ($\rm{M_{\odot}}$) & $2.789\pm 0.139$ & \citet{Kervella19}\\
$M_{\ast}$ ($\rm{M_{\odot}}$) & $2.14\pm 0.11$ & \citet{Gaulme20}\\
$R_{\ast}$ ($\rm{R_{\odot}}$) & $11.2 \pm 0.6$ & This work\\
$R_{\ast}$ ($\rm{R_{\odot}}$) & $9.996\pm 0.851$ & \citet{Ghezzi15}\tablefootmark{c}\\
$R_{\ast}$ ($\rm{R_{\odot}}$) & $11.255\pm 0.718$ & \citet{Ghezzi15}\tablefootmark{b}\\
$R_{\ast}$ ($\rm{R_{\odot}}$) & $10.686\pm 0.534$ & \citet{Kervella19}\\
$R_{\ast}$ ($\rm{R_{\odot}}$) & $10.01\pm 0.21$ & \citet{Gaulme20}\\
Age (Gyr) & $8.64 \pm 0.17$ & This work\\
Age (Gyr) & $0.993\pm 0.298$ & \citet{Ghezzi15}\tablefootmark{c}\\
Status & He-core burning & This work\\
Status & RGB\tablefootmark{d} & \citet{Thygesen12}\\
Status & Red Clump & \citet{Gaulme20}\\
\hline
  \end{tabular}
\tablefoot{\tablefoottext{a}{Spectroscopic analysis;}\tablefoottext{b}{Asteroseismic
analysis;}\tablefoottext{c}{Evolutionary tracks;}\tablefoottext{d}{Red Giant Branch
star.}}
\end{table} 

The basic stellar parameters were determined from a high-resolution
(R=85,000) spectrum of HD~187878 taken with the HERMES spectrograph at the
Mercator telescope with a S/N of $\sim 160$. We followed
the same procedure as in Sect.~\ref{properties_kic} on a wavelength range 
4690 -- 6700~\AA~which provided the best results. The blue spectral range was omitted 
because of the presence of dense molecular CNO lines that cannot be properly 
handled by \textsc{SynthV} \citep{Tsymbal96}. 

Results of abundances of chemical elements are listed in 
Table~\ref{hd_abundances}. HD~187878 has a solar metallicity. 
We note that abundances of C and N deviate from solar values which confirms
the trend of a nitrogen enrichment and a carbon deficiency in most giant
stars as compared with main sequence stars \citep{Kjaergaard82}.

The \textsc{Octave} (Birmingham - Sheffield Hallam) automated pipeline
\citep{Hekker10} was used to determine the large
frequency separation between modes of consecutive order and same degree,
$\Delta\nu$, and the frequency of maximum oscillation power, $\nu_{\rm max}$.
These values were combined in
a grid-based modelling \citep{Hekker13} using $T_{\rm eff}$ and [Fe/H] from the
spectroscopic measurements to determine the stellar mass, radius, age and $\log g$. 
All resulting parameters are shown in Table~\ref{hd_stellar_par}.

\begin{table*}
\caption{Abundances of HD~187878 relative to solar composition.}\label{hd_abundances}
\centering
\begin{tabular}{cccccccccc}
\hline
\noalign{\smallskip}
C & N & O & Na & Mg & Si & Ca & Sc & Ti & V\\
$-0.24_{-0.14}^{+0.10}$ &
$+0.81_{-0.24}^{+0.17}$ &
$+0.10_{-0.59}^{+0.44}$ &
$+0.38_{-0.26}^{+0.23}$ &
$+0.18_{-0.13}^{+0.12}$ &
$-0.30_{-0.24}^{+0.20}$ &
$+0.07_{-0.28}^{+0.26}$ &
$+0.13_{-0.38}^{+0.30}$ &
$+0.13_{-0.12}^{+0.11}$ &
$+0.11_{-0.21}^{+0.19}$ \\
\noalign{\smallskip}
\noalign{\smallskip}
Cr & Mn & Fe & Co & Ni & Y & Ba & Ce & Nd & Sm\\
$+0.06_{-0.15}^{+0.13}$ &
$+0.18_{-0.26}^{+0.23}$ &
$ 0.00_{-0.05}^{+0.05}$ &
$-0.04_{-0.20}^{+0.17}$ &
$-0.04_{-0.14}^{+0.13}$ &
$+0.08_{-0.49}^{+0.41}$ &
$+0.55_{-0.81}^{+0.44}$ &
$+0.18_{-0.68}^{+0.32}$ &
$+0.24_{-0.27}^{+0.22}$ &
$+0.20_{-0.86}^{+0.39}$ \\
\noalign{\smallskip}
\hline
  \end{tabular}
\end{table*} 

\subsection{Companion to HD~187878}\label{orbit_hd}

\begin{figure}
\includegraphics[width=\hsize]{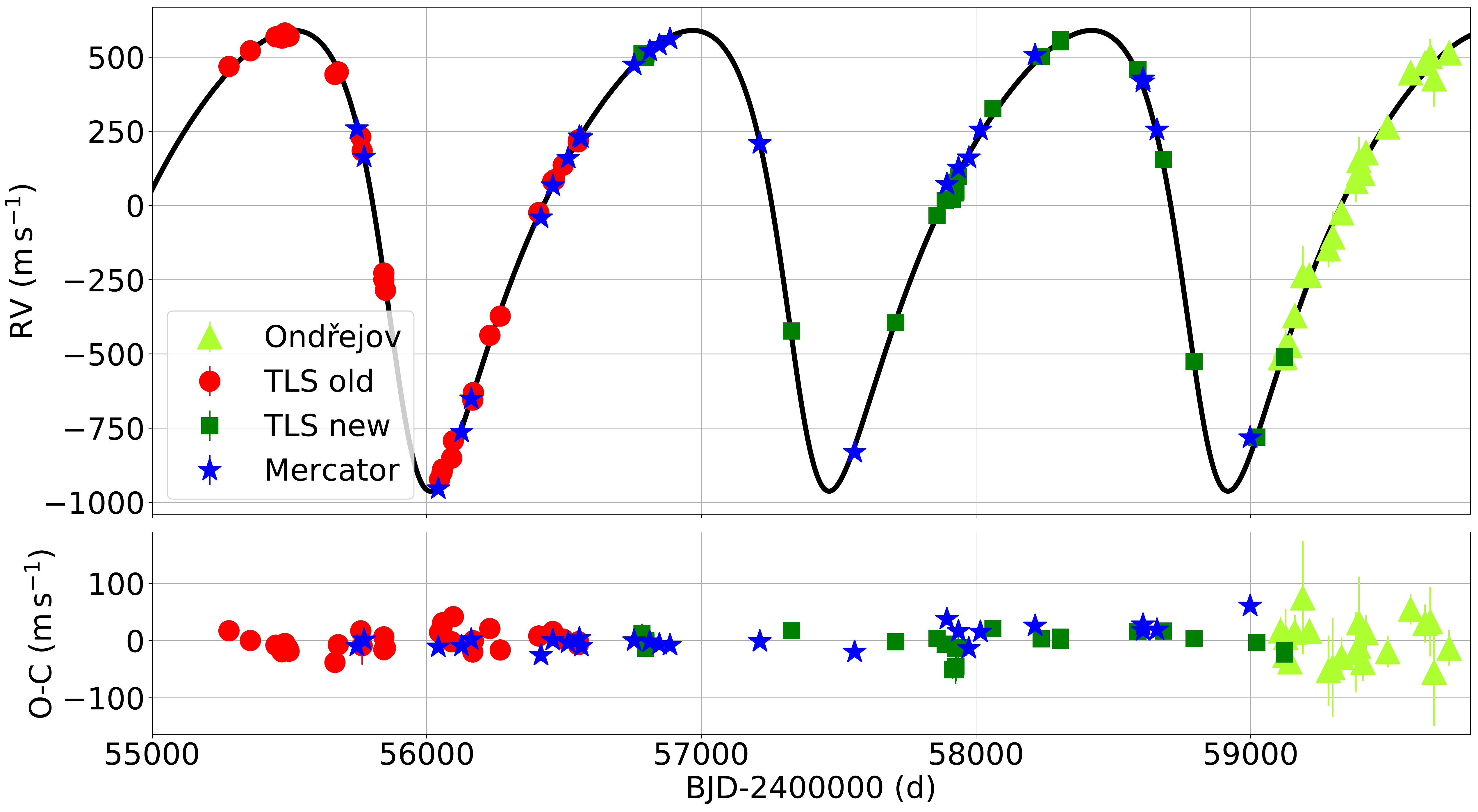}
\caption{RV measurements of HD~187878. Top: Data obtained from March 2010 to 
May 2022 using the coud\'e echelle spectrograph at TLS, Germany, the HERMES 
spectrograph at Mercator, La Palma and the Ond\v{r}ejov Echelle
Spectrograph, Czech Republic. The solid curve represents the Keplerian orbital solution. 
Bottom: RV residuals and error bars after removing the 
orbital solution of HD~187878~B.}
\label{Fig_hd_orbital}
\end{figure}

\begin{figure}
\centering
\includegraphics[width=\hsize]{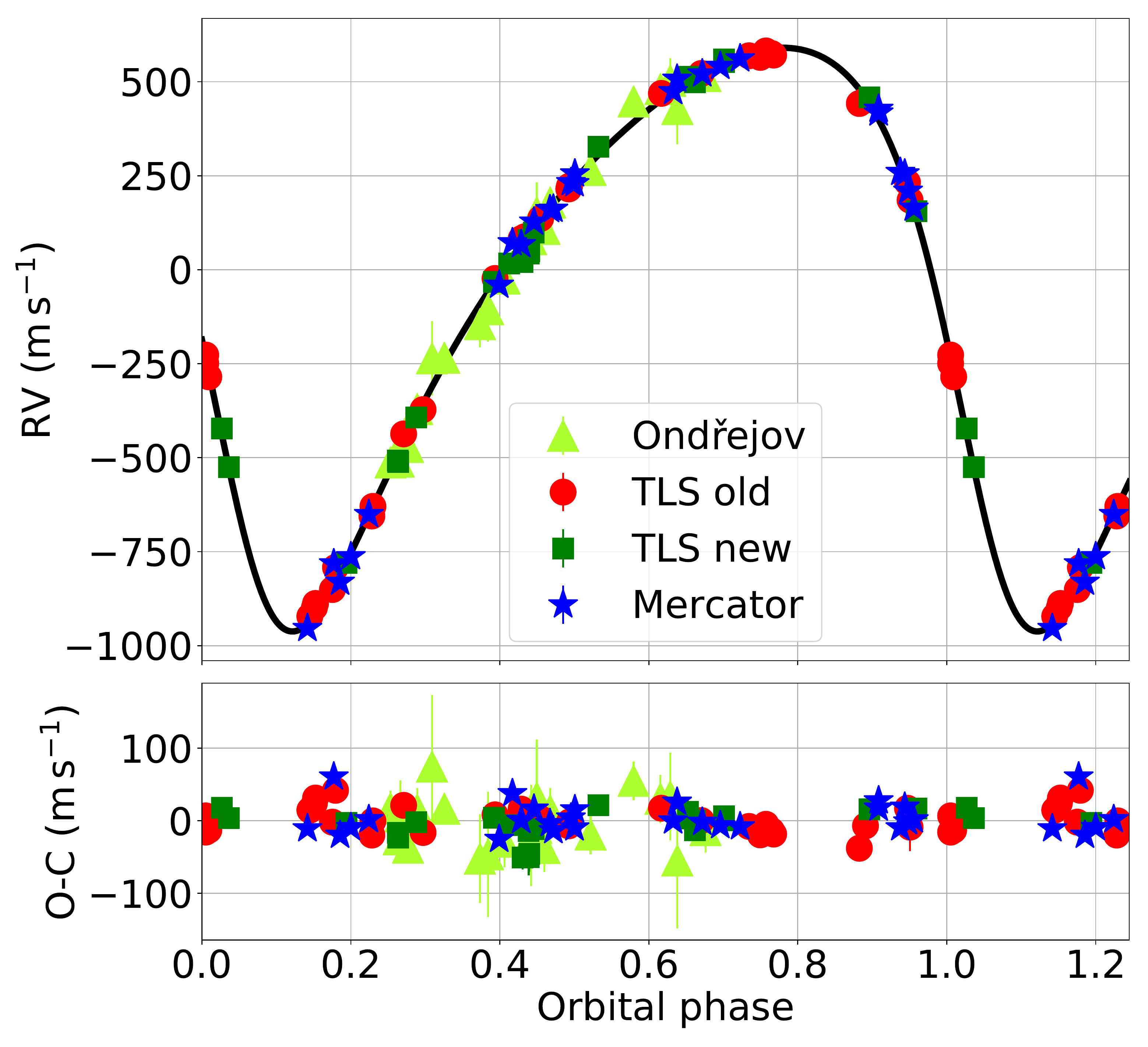}
\caption{Phased RV measurements of HD~187878. Top: Data plotted with corresponding error
bars and phased to the orbital period of 1452.3 d. The Keplerian orbital solution is  
over-plotted with a solid curve. Bottom: RV residuals and error bars 
after removing the orbital solution of HD~187878~B.}
\label{Fig_hd_phased}
\end{figure}

\begin{figure}
\centering
\includegraphics[width=\hsize]{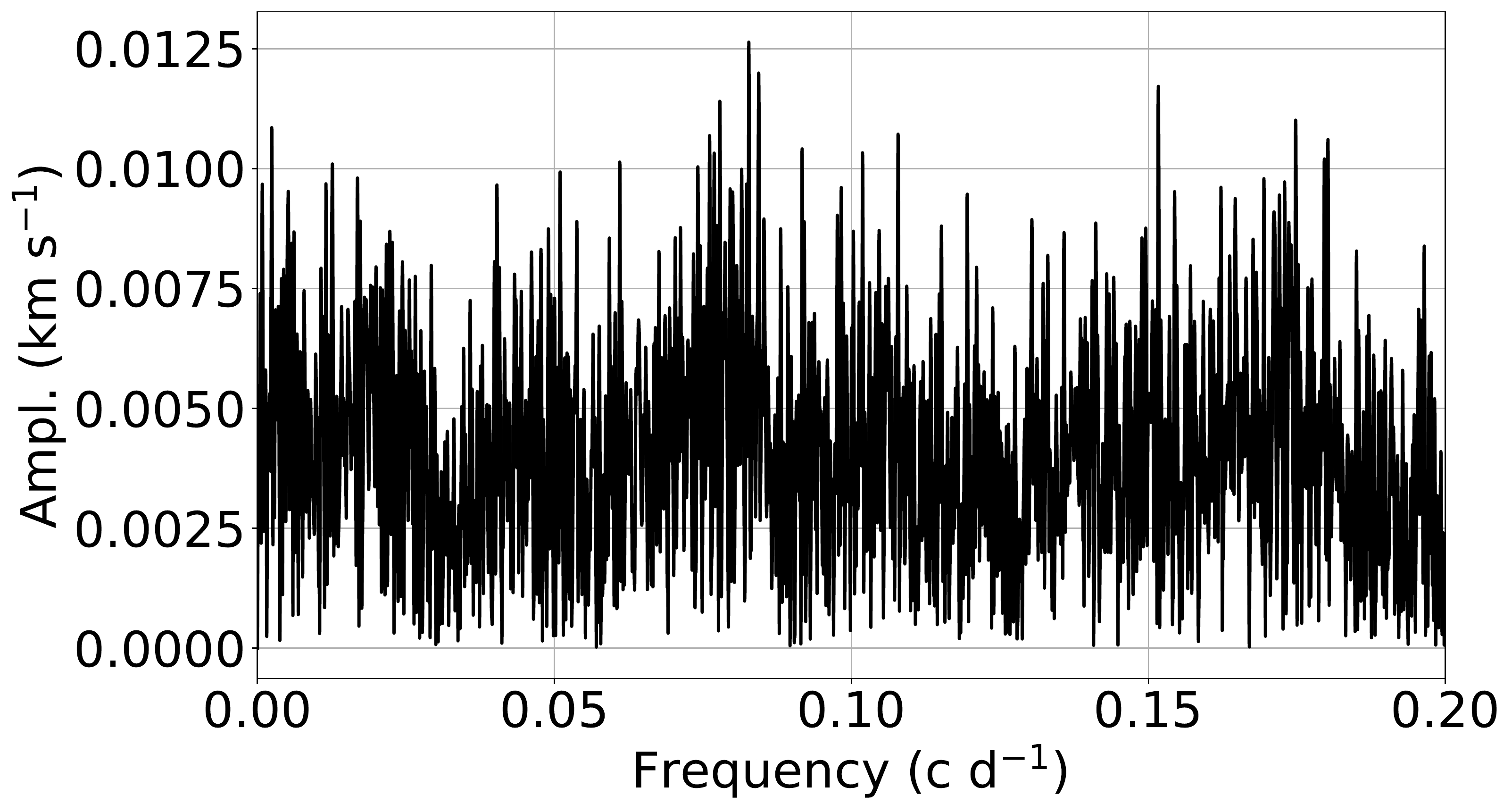}
\caption{Frequency amplitude spectrum of residual RV data of HD~187878 after
removing the Keplerian orbital solution.}
\label{Fig_hd_rv_resid}
\end{figure}

We monitored HD~187878 for a time span of 12 years, during which we
acquired 100 RV measurements (see Table~\ref{hd_rvs} and Fig.~\ref{Fig_hd_orbital}). 
Our RV measurements show changes which could be caused by a low-mass stellar
or a brown dwarf companion. Similarly as for KIC~3526061
we used the code \textsc{pyaneti} \citep{Barragan19} to find 
the orbital solution. We accounted for the RV zero-points between the 
four different data sets with the advantage that data from
different instruments have been taken close in time. 
The inferred parameters are given in Table~\ref{hd_orbit}.
We also used \textsc{FOTEL} \citep{Hadrava04} 
to check independently the orbital solution, and the resulting parameters were 
the same as with \textsc{pyaneti} within 1.5 $\sigma$ uncertainties. 
In Fig.~\ref{Fig_hd_orbital}, we show the RV measurements with the orbital
solution and RV residuals with error bars after removing the orbital solution. 
Finally, the phase-folded RV variations for the orbital solution and the orbital 
fit are shown in Fig.~\ref{Fig_hd_phased}. 
We also searched for additional periods in the residual RV data using the program \textsc{Period04} 
\citep{Lenz05} and found a period of 194 d with a $3.0\,\sigma$ significance.
As this period is very close to a half year and also has a low significance, 
we checked all individual data sets and found that this period is present only in the data from
Ond\v{r}ejov. Therefore, we searched in residual RV data again but with the
Ond\v{r}ejov data excluded, and this time we did not find any significant periods (see
Fig.~\ref{Fig_hd_rv_resid}).

\begin{table}
\caption{Derived parameters of HD~187878~B.}\label{hd_orbit}
\centering
\begin{tabular}{lcc}\hline
Parameter & Value & Unit\\
\hline
\noalign{\smallskip}
\multicolumn{3}{l}{Fitted}\\
\noalign{\smallskip}
Period & $1452.3\pm 0.3$ & d\\
$T_0$ & $2455836.0\pm 0.4$ & BJD\\
$K$ & $776.3\pm 1.6$ & $\rm{m\,s^{-1}}$\\
$e$ & $0.342\pm 0.001$ & \\
$\omega$ & $134.5\pm 0.2$ & deg\\
\noalign{\smallskip}
\multicolumn{3}{l}{Derived}\\
\noalign{\smallskip}
$M\,\sin i$ & $78.4\pm 2.0$ & $\rm{M_{Jup}}$\\
$T_{\,\rm{periastron}}$ & $2455923.1\pm 0.9$ & BJD\\
$a$ & $3.66\pm 0.08$ & AU \\
\noalign{\smallskip}
\multicolumn{3}{l}{Other parameters}\\
\noalign{\smallskip}
$RV_0^{\rm{Ondrejov}}$ & $16.4\pm 7.4$ & $\rm{m\,s^{-1}}$\\
\noalign{\smallskip}
$RV_0^{\rm{TLS\:old}}$ & $-78.2\pm 1.8$ & $\rm{m\,s^{-1}}$\\
\noalign{\smallskip}
$RV_0^{\rm{TLS\:new}}$ & $-42.8\pm 2.3$ & $\rm{m\,s^{-1}}$\\
\noalign{\smallskip}
$RV_0^{\rm{Mercator}}$  & $-18285.3\pm 0.6$ & $\rm{m\,s^{-1}}$\\
${\rm{rms\,}}_{\rm{Total}}$ & 23.3 & $\rm{m\,s^{-1}}$\\
${\rm{rms\,}}_{\rm{Ondrejov}}$ & 35.1 & $\rm{m\,s^{-1}}$\\  
${\rm{rms\,}}_{\rm{TLS\:old}}$ & 17.0 & $\rm{m\,s^{-1}}$\\ 
${\rm{rms\,}}_{\rm{TLS\:new}}$ & 20.5 & $\rm{m\,s^{-1}}$\\ 
${\rm{rms\,}}_{\rm{Mercator}}$ & 19.4 & $\rm{m\,s^{-1}}$\\ 
\noalign{\smallskip}
\multicolumn{3}{l}{Astrometry:}\\
$\quad$Orbital inclination $i$ &$9.8^{+0.4}_{-0.6}$ & deg\\
$\quad$Longitude of asc. node $\Omega$ & $112\pm 2$& deg\\
$\quad$Correlation ($i$, $\Omega$) & -0.35& \\
\noalign{\smallskip}
Mass $M$ & $535^{+44}_{-23}$ & $\rm{M_{Jup}}$\\
\noalign{\smallskip}
Mass $M$ & $0.51^{+0.04}_{-0.02}$ & $\rm{M_{\odot}}$\\
\noalign{\smallskip}
\hline
\end{tabular}
\end{table} 

\subsubsection{Amplitude of stellar oscillations}

The scatter of RV residuals after fitting the companion's orbit is 
23.3$\,\rm{m\,s^{-1}}$. To estimate a velocity amplitude for stellar oscillations
\citep{Kjeldsen95}, we derived the stellar luminosity $L_{\ast}=80.2\pm
9.5\,{\rm{L_{\odot}}}$
using our estimated stellar effective
temperature of $T_{\rm{eff}}=5168\,\rm{K}$ and the stellar radius of
11.2~$\rm{R_{\odot}}$. We used this luminosity and our estimated stellar mass 
of 2.6~$\rm{M_{\odot}}$ to calculate a velocity amplitude of
$v_{osc}=7.2\,\rm{m\,s^{-1}}$. 
We also used Eqs.~7 and 8 from \citet{Kjeldsen95} to predict a velocity amplitude
based on the luminosity amplitude from the \textit{Kepler} light curves, 
and derived $v_{osc}=1.7\,\rm{m\,s^{-1}}$.
The scatter of RV residuals is larger than both velocity amplitude estimates 
and most likely is due to uncertainties in RV measurements. 

\subsubsection{Orbital inclination and mass of a companion}\label{astrometric_hd}

\citet{Kervella19} analysed proper motion anomalies of nearby stars 
to characterise the presence of physical companions of stellar and sub-stellar mass. 
They used the \textit{Hipparcos} catalogue \citep{VanLeeuwen07} and \textit{Gaia}'s 
second data release \citep{Gaia16, Gaia18} to determine the long-term proper motion 
of stars common to the two catalogues. They searched for a proper motion anomaly
by comparing the long-term \textit{Hipparcos}-\textit{Gaia} and short-term 
\textit{Gaia} proper motion vectors of each star, indicating a presence 
of a perturbing secondary object. Later, \citet{Kervella22} used 
the \textit{Gaia} EDR3 catalogue \citep{Gaia21a, Gaia21b} and improved
the accuracy of the detection of proper motion anomalies.

\begin{figure}
\resizebox{\hsize}{!}
{\includegraphics[width=\hsize]{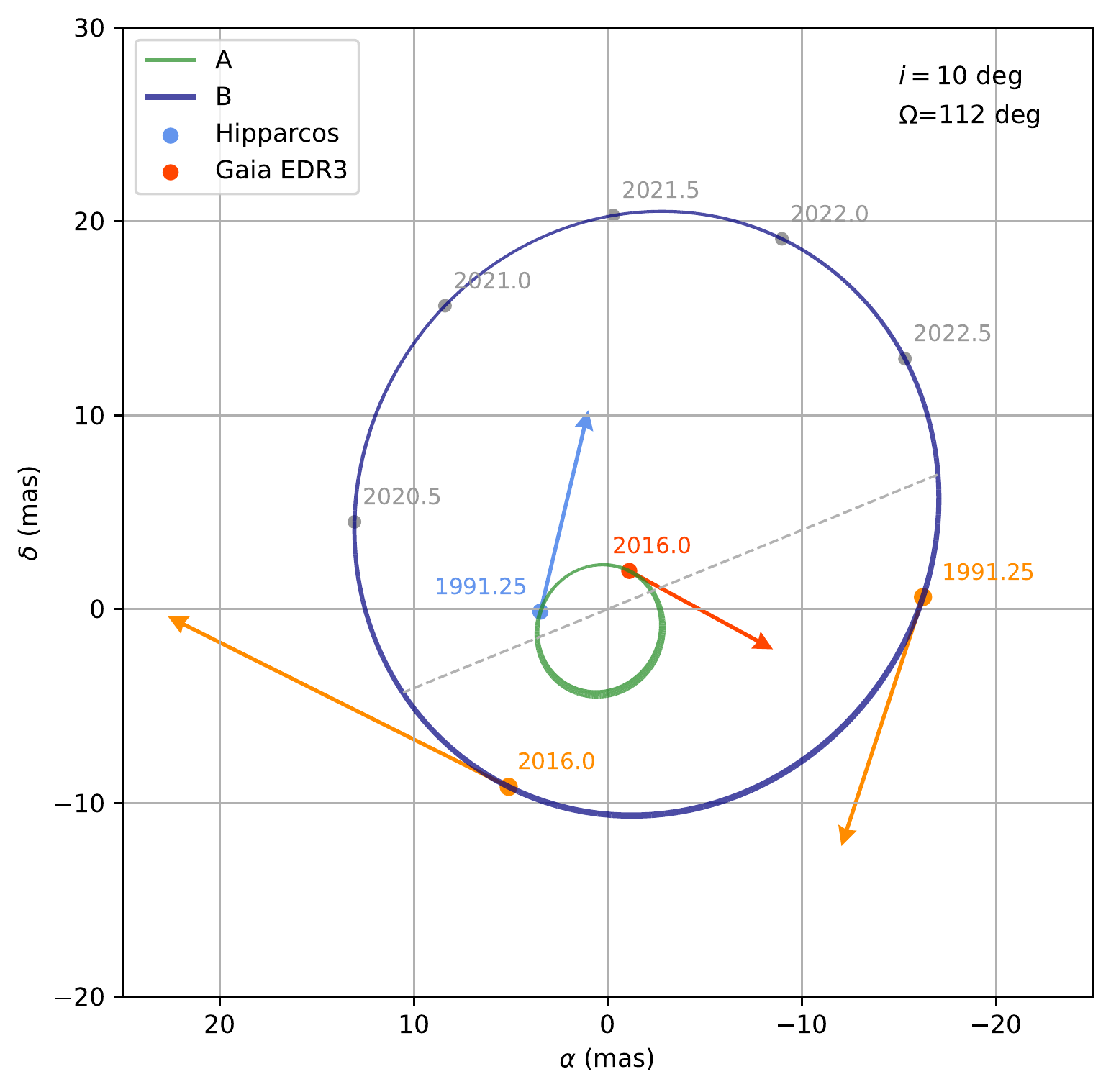}}
\caption{Orbital trajectories of HD~187878 and HD~187878~B around their
centre of mass. A thicker line indicates that the companion is closer to the Earth. 
The orange arrows show the velocity vector of HD~187878~B at the effective 
EDR3 and Hipparcos epochs. Although the Hipparcos proper motion anomaly
vector is not fit, one can see that it is nicely tangent to the orbit.}
\label{Fig_hd_orbit_sky}
\end{figure}

We used a combination of our spectroscopic orbital parameters presented in
Table~\ref{hd_orbit} and \textit{Gaia} EDR3 astrometric proper motion anomaly 
\citep{Kervella22} to derive an orbital inclination and companion's mass.
We also used the Hipparcos proper motion anomaly \citep{VanLeeuwen07} 
to determine that the direction of the orbit is prograde. 
More detailed description of the method can be found in \citet{Kervella20}.   
The inclination is found to be $i=9.8^{+0.4}_{-0.6}$ deg which corresponds to the
stellar companion's mass of $M=535^{+44}_{-23}\,\rm{M_{Jup}}$ or 
$0.51^{+0.04}_{-0.02}\,\rm{M_{\odot}}$ (see Table~\ref{hd_orbit}). 
For this computation, we adopted a stellar 
mass of $2.789\pm 0.139\,\rm{M_{\odot}}$ \citep{Kervella19}, 
and we also took into account the companion's mass
and it's eccentric orbit. The best-fit orbit is
displayed in Fig.~\ref{Fig_hd_orbit_sky}. We note that the companion's mass estimate in the
catalogue of \citet{Kervella22} is comparable to that we obtain through a more refined analysis 
including the RVs.

\subsection{Stellar activity analysis of HD~187878}\label{activity_hd}

\subsubsection{Photometric variations}\label{activityphot_hd}

We analysed the \textit{Kepler} photometry of HD~187878 similarly as for
KIC~3526061. The star was observed by the \textit{Kepler} 
satellite \citep{Borucki10} since May 2009 and for all quarters during the 
main \textit{Kepler} mission. 
The \textit{Kepler} data have similar length as the orbital period of
HD~187878~B. We do not find any long-term trends in the light curve.
There is a feature around 600 Barycentric Kepler
Julian Date (BKJD) which is most likely caused by an improper removal of 
systematic trends when creating PDCSAP light curves. Although the aperture for
HD~187878 which was used to generate PDCSAP light curves contained several 
other stars with \textit{Gaia} magnitudes G=14.1, 14.3, 15.5 and 16.8,
there is no obvious contamination of the flux of HD~187878, which has G=6.9.  

HD~187878 has also been observed in sectors 14 and 15 of the TESS satellite 
\citep{Ricker14}. We used PDCSAP light curves as above for the 
\textit{Kepler} data, and found no long-term variations. 

\subsubsection{Stellar activity indicators}\label{activityindicators_hd}

For the spectral line bisector analysis 
we proceeded as in Sect.~\ref{activitybis_kic} for the Mercator data.
We measured the BVS of the spectral profile using the difference of the
average bisector values between flux levels of 0.1 -- 0.3 and 0.7 -- 0.9
of the continuum value. A search for periods in the BVS variations 
using the program \textsc{Period04} \citep{Lenz05} did not find any
significant signal. The most significant frequency was found at the period of
328.5~d with a low 2.3 $\sigma$ significance.

Similarly as in Sect.~\ref{activityfwhm_kic} we searched for periods in
the FWHM variations and did not find any
significant signal. The most significant frequency was found at the period of
556.1~d with a low 2.7 $\sigma$ significance.

We also analysed the chromospheric activity of HD~187878.
In the Mercator data, first we removed telluric lines in the \ion{Ca}{ii} 8498
\AA~line. Then we measured the EQW at the central part of the 
\ion{Ca}{ii} 8498 \AA~line from 8497.45 to 8498.75 \AA, of the
\ion{Ca}{ii} 8542 \AA~line in the range $\pm 0.65$~\AA~from the line centre, 
and of the \ion{Ca}{ii} 8662 \AA~line in two different ranges, 8661.6 to 8662.78 \AA~and 
8661.1-8663.28 \AA. For each spectrum, we made an average EQW based on the three \ion{Ca}{ii}
lines and four different measurements, and finally searched for periods 
using the program \textsc{Period04} \citep{Lenz05}. The most significant frequency 
was found at the period of 4065~d with a low significance of 2.8 $\sigma$.
To summarise, we did not find any significant periods in any of the stellar activity 
indicators which were investigated. 

\section{Conclusions}\label{Conclusions}

We used precise stellar RV measurements of the intermediate-mass red giant 
branch star KIC~3526061 and the relatively massive, evolved red giant branch 
star HD~187878 to discover variations which we attribute to the presence of
companions. We conclude that RV variations of KIC~3526061 are caused by
a long-period eccentric companion that is very likely a brown dwarf, 
and that RV variations of HD~187878 are due to a stellar companion. 

For HD~187878~B we used a combination of spectroscopic orbital parameters 
and \textit{Gaia} EDR3 astrometric proper motion anomaly \citep{Kervella22} to
derive an orbital inclination and companion's mass. We also used the Hipparcos 
proper motion anomaly \citep{VanLeeuwen07} to determine that the direction of 
the orbit is prograde. The inclination is found to be $i=9.8^{+0.4}_{-0.6}$ deg 
which corresponds to the companion's mass in the stellar regime of 
$535^{+44}_{-23}\,\rm{M_{Jup}}$ or $0.51^{+0.04}_{-0.02}\,\rm{M_{\odot}}$.

KIC~3526061~b has a minimum mass $18.15\pm 0.44\,\rm{M_{Jupiter}}$, but
the unknown orbital inclination leaves an open question on the real nature 
of the companion. Under the assumption that inclination angles are randomly distributed 
on the sky, there is only a 2.6 \% chance for the companion mass to be
greater than $80\,\rm{M_{Jupiter}}$, which has been thought as a dividing
line between sub-stellar and stellar objects \citep{Burrows01, Hatzes15b,
Chen17}. This means that most likely KIC~3526061~b is a sub-stellar object, 
either a brown dwarf or a giant planet.

The orbital period $3552_{-135}^{+158}$ d and orbital eccentricity $0.85\pm
0.01$ makes KIC~3526061~b a unique sub-stellar companion among those orbiting giant
stars. It has the second largest eccentricity of sub-stellar companions 
orbiting giant stars. The origin of such a highly eccentric orbit is linked with 
the formation and dynamical history of KIC~3526061~b. 
Based on the population-level studies \citep{Ma14,Nielsen19,Wagner19,Bowler20}
it seems more likely that KIC~3526061~b was formed via a gravitational instability. 
However, it is not excluded that KIC~3526061~b formed via a core accretion and 
developed such a large eccentricity through dynamical interactions with other
companions. In addition, it is not clear whether population-level eccentricities 
of sub-stellar companions evolve over time, or they are established at young
ages \citep{Bowler20}. KIC~3526061 as the most evolved system found having a
sub-stellar companion with such a large eccentricity and wide separation  
might provide a probe of the dynamical evolution of such systems over time. 

In a spectral analysis of KIC~3526061 we encountered a large dependency between 
the projected rotational velocity $v\,\sin i$ and the macroturbulent velocity 
$v_{\rm{macro}}$. This is not surprising since there is a tradeoff between $v\,\sin i$
and $v_{\rm{macro}}$ and a slight decrease in $v\,\sin i$ can be compensated by an 
increase in $v_{\rm{macro}}$. We analysed bisector velocity spans, a
full-width at half maximum of spectral lines and equivalent width of the 
\ion{Ca}{ii} triplet lines and found stellar activity variations on a
similar time scale of about 183 days in all three quantities. 
If these variations would be related to a stellar rotation period of 183
days, then the projected rotational velocity would be $v\,\sin i=1.6\,\rm{km\,s^{-1}}$. 
However, we should be cautious as the signal is weak and is about half of the year 
period. It is possible that the variations are due to changes of the instrumental
profile, which is specially true of the McDonald data since there are moving
components in the Tull Spectrograph such as the echelle grating and prisms
and they are changed for different setups according to the observer's needs. 

The search for potential transits of KIC~3526061~b is complicated by solar-like 
oscillations of KIC~3526061 which are larger than an expected transit depth
and their variations have a similar time scale as the expected 
transit duration.
In addition, unfortunately we do not have a good knowledge about potential transit 
times due to a large uncertainty in the orbital period. To refine the orbital solution 
more RV measurements from stable spectrographs would be needed, and particularly 
observations around the time of the predicted periastron passage.
Finally, given such a large orbital period of KIC~3526061~b it shows the
importance of having long enough time series of observations in order to 
understand the occurrence of companions moving on long-period orbits.

\begin{acknowledgements}
Based on observations made with NASA\textquoteright s Discovery mission
\textit{Kepler} and with the HERMES spectrograph,
installed at the Mercator telescope, operated on the island of La Palma by the Flemish Community,
at the Spanish Observatorio del Roque de los Muchachos of the Instituto de
Astrof\'isica de Canarias and supported by the Research Foundation - Flanders (FWO), Belgium, the
Research Council of KU Leuven, Belgium, the Fonds National de la Recherche Scientific 
(F.R.S.-FNRS), Belgium, the Royal Observatory of Belgium, the Observatoire de Gen\`eve, Switzerland
and the Th\"uringer Landessternwarte Tautenburg, Germany.
The data presented here have been taken using the 2-m Alfred Jensch 
Telescope of the Th\"{u}ringer Landessternwarte Tautenburg, the 2.7-m Harlan J. Smith 
Telescope at the McDonald Observatory in Texas and at the 2-m Perek telescope at the
Astronomical Institute of the Czech Academy of Sciences in Ond\v{r}ejov. We are very grateful 
to the personnel of these facilities for their dedication and support during
our observations. We thank our anonymous referee for giving us
many useful suggestions for the improvement of the manuscript.
This research has made use of the electronic bibliography maintained 
by NASA-ADS system and the SIMBAD data base, operated at CDS, Strasbourg,
France. MK acknowledges the support from ESA-PRODEX PEA4000127913. RK, MS
and PKa acknowledge the support by Inter-transfer grant no LTT-20015.  
We acknowledge the use of public TESS data from pipelines at the TESS Science 
Office and at the TESS Science Processing Operations Center. 
This work has made use of data from the European Space Agency (ESA) mission
Gaia (https://www.cosmos.esa.int/gaia), processed by the Gaia Data
Processing and Analysis Consortium (DPAC, 
https://www.cosmos.esa.int/web/gaia/dpac/consortium). Funding for the DPAC
has been provided by national institutions, in particular the institutions 
participating in the Gaia Multilateral Agreement.
This research made use of Lightkurve, a Python package for Kepler and TESS 
data analysis, and Astropy (http://www.astropy.org), a community-developed 
core Python package for Astronomy. 

\end{acknowledgements}

\bibliographystyle{aa} 
\bibliography{references} 
\end{document}